\documentclass[aps,prl,twocolumn,showpacs,superscriptaddress,groupedaddress]{revtex4-2}  
\usepackage[T1]{fontenc}
\usepackage{lmodern} 
\usepackage{graphicx}  
\usepackage{dcolumn}   
\usepackage{bm}        
\usepackage{amssymb}   
\usepackage{amsmath}
\usepackage{bbold}
\usepackage{array}
\usepackage{makecell}
\usepackage{braket}
\usepackage{physics}
\usepackage{titlesec} 
\usepackage[colorlinks,citecolor=blue,urlcolor=blue,hypertexnames=true]{hyperref}
\usepackage{subfigure}

\usepackage[usenames,dvipsnames,svgnames,table]{xcolor} 
\allowdisplaybreaks

\newcommand{\be}{\begin{equation}}
\newcommand{\ee}{\end{equation}}
\newcommand{\bea}{\begin{eqnarray}}
\newcommand{\eea}{\end{eqnarray}}
\newcommand{\bml}{\begin{subequations}}
\newcommand{\eml}{\end{subequations}}
\newcommand{\bfig}{\begin{figure}}
\newcommand{\efig}{\end{figure}}

\newcommand{\bmat}{\begin{pmatrix}}
\newcommand{\emat}{\end{pmatrix}}

\begin{document}

\widetext


\title{\textcolor{Sepia}{\textbf{\Large Four-mode squeezed states in de Sitter space: A study with two field interacting quantum system}}}

\author{{Sayantan Choudhury ${}^{1,2}$},\thanks{{\it Corresponding author,}\\
	{{ E-mail:sayantan\_ccsp@sgtuniversity.org, sayantan.choudhury@icts.res.in,  sayanphysicsisi@gmail.com}}}~ Sudhakar Panda ${}^{3,4}$, Nilesh Pandey ${}^{5}$ and  Abhishek Roy ${}^{6}$\vspace{.2cm}} 
	
\affiliation{${}^{1}$Centre For Cosmology and Science Popularization (CCSP),
SGT University, Gurugram, Delhi- NCR, Haryana- 122505, India,}
\affiliation{${}^{2}$International Centre for Theoretical Sciences, Tata Institute of Fundamental Research (ICTS-TIFR), Shivakote, Bengaluru 560089, India,}
\affiliation{${}^{3}$National Institute of Science Education and Research, Jatni, Bhubaneswar, Odisha - 752050, India,}
\affiliation{${}^{4}$Homi Bhabha National Institute, Training School Complex, Anushakti Nagar, Mumbai -
400085, India,}
\affiliation{${}^{5}$Department of Applied Physics, Delhi Technological University, Delhi-110042, India,}
\affiliation{${}^{6}$Department of Physics, Indian Institute of Technology Jodhpur,Karwar, Jodhpur - 342037, India.}

\begin{abstract}
    In this paper we study the application of four-mode squeezed states in the cosmological context,  studying two weakly coupled scalar fields in the planar patch of the de Sitter space.  We construct the four-mode squeezed state formalism and connect this concept with the Hamiltonian of the two coupled inverted harmonic oscillators having a time-dependent effective frequency in the planar patch of the de Sitter space.  Further, the corresponding evolution operator for the quantum Euclidean vacuum state has been constructed,  which captures its dynamics.  Using the Heisenberg picture coupled differential equations describing the time evolution for all squeezing parameters (amplitude,  phase and angle) have been obtained, for the weakly coupled two scalar field model.  With the help of these evolutions for the coupled system, we simulate the dynamics of the squeezing parameters in terms of conformal time.  From our analysis, we observe interesting dynamics,  which helps us to explore various underlying physical implications of the weakly coupled two scalar field system in the planar patch of the de Sitter cosmological background.

\end{abstract}

\pacs{}
\maketitle

\section{\textcolor{Sepia}{\textbf{ \Large Introduction}}} \label{sec:introduction}
The squeezed states in quantum optics have emerged as non-classical states of light which are a consequence of Heisenberg's uncertainty relations. When the uncertainty in one conjugate variable is below the symmetric limit,  compared to the other conjugate variable,  without affecting the Heisenberg's uncertainty relations,  then the state obtained is called squeezed state.  For review on the fundamentals of squeezed states see \cite{Ph:31,ssol,book,sql,30yrof,RosasOrtiz2019CoherentAS,Schumaker:1985zz,Garcia-Chung:2020gxy}. These states have been used in the field of quantum optics \cite{PhysRevA.31.3068,Zubairy_2005,Furusawa:07,ncsqo,ZELAYA20183369} for many experimental purposes.  From an application perspective the squeezed states of light are being used for various applications in quantum computing \cite{PhysRevLett.97.110501,PhysRevA.103.062405} and quantum cryptography \cite{PhysRevA.61.022309} because these non-classical states of light can perform as an elementary resource in quantum information processing for the continuous variable systems. These states are also now being used in gravitational wave physics \cite{PhysRevLett.97.011101,Chua_2014,Choudhury:2013jya} to enhance the sensitivity of gravitational wave detectors \cite{Darsow-Fromm:21,qmetr,Barsotti:2018hvm,Schnabel:2016gdi},  such as the LIGO \cite{enhans} and to improve measurement techniques in quantum metrology \cite{Vahlbruch_2007,PhysRevLett.104.103602,advinqm,doi:10.1080/00107510802091298,Xu:19,atomchp,Choudhury:2020dpf}.
For a more broad class of application of squeezed states see \cite{1981PhRvD..23.1693C,RevModPhys.77.513,contivar,braunstein2005quantum,PhysRevLett.117.110801,RevModPhys.58.1001,Adhikari:2021ked,Ando:2020kdz,Martin:2021znx,Bhargava:2020fhl,Choudhury:2020hil,Adhikari:2021pvv,Choudhury:2021brg}.


From the cosmological point of view,  the use of the formalism of squeezed states was introduced in one of the early works by Grishchuk and Sidorov \cite{Grishchuk:1990cm,Grishchuk:1991cm} on the inflationary cosmology where they analysed the features of relic gravitons and phenomena such as particle creation and black-hole evaporation using the two-mode squeezed state formalism.  They showed that the amplification of quantum fluctuations into macroscopic perturbations which occurs during cosmic inflation is a process of quantum squeezing and in the cosmological scenario they describe primordial density perturbations,  amplified by gravitational instability from the quantum vacuum fluctuations. Another important work by Andreas Albrecht et al. \cite{Albrecht:1992kf} have used the two-mode squeezed state formalism (for a single field) to understand the inflationary cosmology and the amplification process of quantum fluctuations during the inflationary epoch.  Another more recent work \cite{Colas2021FourmodeSS} discussed on four-mode squeezed states for two quantum systems using the symplectic group theory and its Lie algebra.  These works have motivated us to explore both the theoretical and numerical features of two coupled scalar fields in the planar patch of the de Sitter space by developing the formalism and constructing the corresponding four-mode squeezed operator.  For more vast applications of squeezed state formalism in High energy physics and in cosmology see   \cite{Hasebe:2019ibg,Choudhury:2011jt,Choudhury:2012yh,Choudhury:2011sq,Choudhury:2013zna,Choudhury:2015hvr,Bhattacharyya:2020kgu,Choudhury:2017cos,Akhtar:2019qdn,Choudhury:2016pfr,Choudhury:2016cso,Bhattacharyya:2020rpy,Choudhury:2017bou,Einhorn:2003xb,Choudhury:2017qyl,Baumann:2014nda,Grain:2019vnq,Grishchuk:1992tw,Bhargava:2020fhl,Choudhury:2020hil,Adhikari:2021pvv,Choudhury:2021brg,Martin:2021qkg}.



The structure of the paper can be broadly divided into two parts, {section \ref{sec:E2}} and {section \ref{LCDE}}.

In the {section \ref{sec:E2}},  we start our analysis by considering two massive scalar fields in the planar patch of de Sitter space with $K$ as weak coupling constant between the fields.  In {subsection \ref{motiv}}, we will quantize the modes of the two coupled scalar fields.  Here for simplicity we will neglect the ``back action'' of the scalar fields to the de Sitter background geometry.  We will find the quantized position and momentum variables for the two coupled scalar fields on the planar patch of the de Sitter space and then compute the quantized version of the Hamiltonian,  for our case.  In {subsection \ref{Inverte}}, we will briefly present the connection between the two coupled inverted quantum harmonic oscillator system and four mode squeezed state formalism constructed for our model under consideration.  In {subsection \ref{fmss1}}, we will construct four mode squeezed state operator which will be useful for understanding the cosmological implications for two interacting  scalar fields and also for other systems which can be explained in terms of two coupled inverted quantum harmonic oscillators.

In the {section \ref{LCDE}},  we will define the time evolution operator for the four mode squeezed states,  which we will use to calculate the time dependent (Heisenberg picture) annihilation and creation operators for two coupled scalar fields in the planar patch of the de Sitter space.  
In {subsection \ref{eqdiff1}}, we use the Heisenberg equation of motion to calculate the coupled differential equation for the mode functions of the two coupled scalar fields in the planar patch of the de Sitter space.  We will also calculate the position and momentum operators in Heisenberg representation for this model.  After that we give the expression for mode functions by using the set of coupled differential equation for the mode functions.
We give the expression for the squeezing parameters,  $R_{1,\bf{k}}$, $\Theta_{1,\bf{k}}$, $\Phi_{1,\bf{k}}$, $R_{2,\bf{k}}$, $\Theta_{2,\bf{k}}$ and $\Phi_{2,\bf{k}}$ which governs the evolution of the quantum state for two coupled scalar fields in the planar patch of de Sitter space.

\section{\textcolor{Sepia}{\textbf{ \Large Four-mode Squeezed State Formalism for two field interacting model}}}
\label{sec:E2}
In the following section, our prime objective is to construct a formalism for two scalar fields $\phi_1$ and $\phi_2$,  in the planar patch of de Sitter space,  which are  weakly interacting with each other through a coupling strength $K$.  The action for the two scalar fields contain a usual gravitational part with $R$ as Ricci scalar,  a matter source term $T$,  it also contains kinetic term and potential term from both the fields.  


The corresponding action for two coupled interacting scalar fields \cite{PhysRevD.42.3413} in the planar patch of de Sitter space can be written as:
\begin{multline}
      S=\int d^{3}x dt \sqrt{-g} \bigg[-\frac{R}{16 \pi G} + T + \frac{1}{2}\left( \dot{\phi_{1}^{2}}- m_{1}^{2} \phi_{1}^{2}\right) \\ + \frac{1}{2}\left( \dot{\phi_{2}^{2}}- m_{2}^{2} \phi_{2}^{2}\right)-12KR\phi_{1}\phi_{2} \bigg]
      \label{acti}
\end{multline}
Here,  the corresponding de Sitter metric in the planar patch is described by the following line element:
\begin{equation}
ds^{2}=-dt^{2}+a^{2}(t)d{\bf x}^{2}\quad\quad {\rm with}\quad a(t)=\exp(Ht).
\end{equation}
Here $``a$'' is the scale factor which is a function of time and $H$ is the Hubble constant.
Also here $m_1$ is the mass of field $\phi_1$ and $m_2$ is the mass of field $\phi_2$.  Due to having the homogeneity and isotropy in the spatially flat FLRW background having planar de Sitter solution both the fields are only functions of time and there is no space dependence.  For this reason there is no kinetic term appearing which involve the space derivatives.  Before moving further we introduce some conditions which will help us in simplifying the calculations.
We define the integral over the matter source term to be proportional to the potential function $V(a)$ and is given by:
\begin{equation}
 \int d^{3} x \sqrt{-g}~T=\frac{1}{2} V(a)
 \end{equation}\\ 
We will set the Planck length $l_{p}=1$ for the rest of the computation and we also introduce some dimensionless field  variables,  which are given by: 
\begin{equation}
\begin{aligned}
&\mu_{1}(t)= \phi_{1}(t) a^{3 / 2}(t)=\exp(3Ht/2) \phi_{1}(t),  \\
&\mu_{2}(t)= \phi_{2}(t) a^{3 / 2}(t)=\exp(3Ht/2) \phi_{2}(t). \\ 
\end{aligned}
\end{equation}
After doing all this manipulations and using the field redefinition the Lagrangian for two coupled scalar fields in the planar patch of de Sitter space can be recast in the following simplified form:
\begin{eqnarray}
L=-\frac{a \dot{a}^{2}}{2}+\frac{V(a)}{2}+l_{1}+l_{2}+l_{3}\label{eq3c}
\end{eqnarray}
Where, $l_1$, $l_2$ and $l_3$ are given by the following expressions:
\begin{eqnarray}
&&l_{1}=\frac{\dot{\mu}_{1}^{2}}{2}-\frac{1}{2} \mu_{1}^{2} m_{1}^{2}-\frac{3}{2} \frac{\dot{a}}{a} \dot{\mu_{1}} \mu_{1} +\frac{9}{8}\left(\frac{\dot{a}}{a}\right)^{2} \mu_{1}^{2},\\
&&l_{2}=\frac{\dot{\mu}_{2}^{2}}{2}-\frac{1}{2} \mu_{2}^{2} m_{2}^{2}-\frac{3}{2} \frac{\dot{a}}{a} \dot{\mu_{2}} \mu_{2}+\frac{9}{8}\left(\frac{\dot{a}}{a}\right)^{2} \mu_{2}^{2},\\
&&l_{3}=K\left[\left(\frac{\dot{a}}{a}\right)^{2} \mu_{1} \mu_{2}-\frac{1}{2} \frac{\dot{a}}{a}\left(\dot{\mu_{1}} \mu_{2}+\mu_{1} \dot{\mu_{2}}\right)\right].
\end{eqnarray}
With the use of Euler Lagrange equations we get the two equation of motion each for $\mu_1$ and $\mu_2$ fields are given by:
\begin{eqnarray}
&&\ddot{\mu_{1}}+\left[m_{1}^{2}-\frac{3}{2} \frac{\ddot{a}}{a}-\frac{3}{4}\left(\frac{\dot{a}}{a}\right)^{2}\right] \mu_{1} \nonumber\\
&&\quad\quad\quad-\frac{K}{2}\left[\frac{\ddot{a}}{a}+\left(\frac{\ddot{a}}{a}\right)^{2}\right] \mu_{2}=0,\\
&&\ddot{\mu}_{2}+\left[m_{2}^{2}-\frac{3}{2} \frac{\ddot{a}}{a}-\frac{3}{4}\left(\frac{\dot{a}}{a}\right)^{2}\right] \mu_{2} \nonumber\\
&&\quad\quad\quad-\frac{K}{2}\left[\frac{\ddot{a}}{a}+\left(\frac{\dot{a}}{a}\right)^{2}\right] \mu_{1}=0.
\end{eqnarray}
Here we can clearly notice that the above two equations are coupled differential equation of motion due to having the interaction in the original theory.
  
Now we can construct the Hamiltonian for our theory from the Lagrangian given in Eq\eqref{eq3c}. The Hamiltonian for two interacting scalar fields in the planar patch of de Sitter background is given by:
\begin{equation}
     H=-\frac{\pi^{2}}{2 M} -\frac{V(a)}{2} + \frac{1}{2}\left(\pi_{1}^{2}+m_{1}^{2} v_{1}^{2}\right)
+\frac{1}{2}\left(\pi_{2}^{2}+m_{2}^{2} v_{2}^{2}\right)
\end{equation}
Where,  we define:

\begin{align}
M&=a+\frac{K}{a^{2}}\left[\mu_{1} \mu_{2}+\frac{K}{4}\left(\mu_{1}^{2}+\mu_{2}^{2}\right)\right] \\
\pi&=p_{a}+\frac{p_{1}}{2 a}\left(3 \mu_{1}+k \mu_{2}\right)+\frac{p_{2}}{2 a}\left(3 \mu_{2}+k \mu_{1}\right)
\end{align}
Here we introduce,  $p_{a}$ : Canonically conjugate momenta of scale factor $a$, 
$\pi_{1}$, $\pi_{2}$ : Canonically conjugate momenta of redefined fields $\mu_{1}$ and $\mu_{2}$ respectively.

\subsection{\textcolor{Sepia}{\textbf{Quantizing the Hamiltonian}}} \label{motiv}

We will be quantizing the fields $\mu_{1}$ and $\mu_{2}$, and will be treating gravity classically in this computation.  In our analysis the back reaction from the fields is neglected.  For this reason we expand $\frac{\pi^{2}}{2 M}$ in a series and retain only the terms $\frac{\pi_{a}^{2}}{a}$ (which governs together with the potential $V(a)$ for a which is a semi-classical behaviour of the background geometry).

With these condition defined above we get the approximated form of the Hamiltonian as follows:
\begin{equation}
\begin{aligned}
H &\sim \frac{1}{2}\left(\pi_{1}^{2}+v_{1}^{2} m_{1}^{2}\right)+\frac{1}{2}\left(\pi_{2}^{2}+v_{2}^{2} m_{2}^{2}\right)\\
&-\frac{p_{a}^{2}}{2 a^{2}}\left[3\left(v_{1} \pi_{1}+v_{2} \pi_{2}\right)\right.\left.\left.+ K\left(v_{1} \pi_{2}+v_{2}\pi_{1}\right\}\right)\right]\\
&+\frac{K p_{a}^{2}}{2 a^{4}}\left(v_{1} v_{2}+\frac{k}{4}\left(v_{1}^{2}+v_{2}^{2}\right)\right)
\end{aligned}
\end{equation}
where we define $p_{a}\sim - a\dot{a}$.

Now we will quantize this Hamiltonian.  For this purpose we promote the fields to operators and take the Fourier decomposition.  We use the following ansatz for Fourier decomposition:

\begin{equation}
  \begin{aligned}
    \hat{v}_{1} &=\int \frac{d^{3} k}{(2 \pi)^{3}} \hat{v}_{\textbf{1,k}} e^{i \textbf{k} \cdot \textbf{x}}, \\
    \hat{\pi}_{1} &=\int \frac{d^{3} k}{(2 \pi)^{3}} \hat{\pi}_{\textbf{1,k}} e^{i \textbf{k} \cdot \textbf{x}},\\
    \hat{v}_{2} &=\int \frac{d^{3} k}{(2 \pi)^{3}} \hat{v}_{\textbf{2,k}} e^{i \textbf{k} \cdot \textbf{x}}, \\
    \hat{\pi}_{2} &=\int \frac{d^{3} k}{(2 \pi)^{3}} \hat{\pi}_{\textbf{2,k}} e^{i \textbf{k} \cdot \textbf{x}}.
    \end{aligned}  
\end{equation}

For our purpose we will be working in the Schr\"{o}dinger picture, where the operators $\hat{v}_{1,\textbf{k}}$, $\hat{\pi}_{1,\textbf{k}}$, $\hat{v}_{2,\textbf{k}}$ and $\hat{\pi}_{2,\textbf{k}}$ are fixed at an initial time.  We define modes and the associated canonically conjugate momenta for the two fields with initial frequency equal to $k$ which,  suitably normalized,  give us:
 
\begin{equation}
\begin{aligned}
&\hat{v}_{1,\textbf{k}}=\frac{1}{\sqrt{2 k}_{1}}\left(b_{1,\textbf{k}}+b_{1,\textbf{k}}^{\dagger}\right),~~~~~~~\hat{\pi}_{1,\textbf{k}}=i \sqrt{\frac{k_{1}}{2}}\left(b_{1,\textbf{k}}^{\dagger}-b_{1,\textbf{k}}\right) \\
&\hat{v}_{2,\textbf{k}}=\frac{1}{\sqrt{2 k}_{2}}\left(b_{2,\textbf{k}}+b_{2,\textbf{k}}^{\dagger}\right),~~~~~~~\hat{\pi}_{2,\textbf{k}}=i \sqrt{\frac{k_{2}}{2}}\left(b_{2,\textbf{k}}^{\dagger}-b_{2,\textbf{k}}\right)
\end{aligned}
\end{equation}

The Four-mode Hamiltonian operator after quantization for the two scalar fields interacting with each other via coupling constant $K$ in the planar patch of de Sitter space can be written in the following simple form:
\begin{equation}\label{hamilto}
  \begin{aligned}
    H(\tau)&=\left(l_{1}(\tau) b_{1 ,\mathbf{-k}} b_{1, \mathbf{k}}+l_{1}^{*}(\tau) b_{1, \mathbf{k}}^{\dagger} b_{1 ,\mathbf{-k}}^{\dagger}\right)\\
    &+\left(l_{2}(\tau), b_{2, \mathbf{-k}} b_{2 ,\mathbf{k}}+l_{2}^{*}(\tau) b_{2, \mathbf{k}}^{\dagger} b_{2, \mathbf{-k}}^{\dagger}\right)\\
    &+\left\{\omega_{1}(\tau)\left(b^{\dagger}_{1 ,\mathbf{-k}}b_{1 ,\mathbf{k}}+b^{\dagger}_{1, \mathbf{k}} b_{1 ,\mathbf{-k}}\right)\right.\\
    &+\omega_{2}(\tau)\left(b_{2, \mathbf{-k}}^{\dagger}b_{2 ,\mathbf{k}}+b_{2,\mathbf{k}}^{\dagger} b_{2 ,\mathbf{-k}}\right)\\
    &+g_{1}(\tau)\left(b_{1 ,\mathbf{-k}} b_{2 ,\mathbf{k}}+b_{1 ,\mathbf{k}} b_{2, \mathbf{-k}}\right)\\
    &+g_{1}^{*}(\tau)\left(b_{2 ,\mathbf{k}}^{\dagger} b_{1, \mathbf{-k}}^{\dagger}+ b_{2, \mathbf{-k}}^{\dagger} b_{1, \mathbf{k}}\right)\\
    &+g_{2}(\tau)\left( b^{\dagger}_{1 ,\mathbf{k}} b_{2 ,\mathbf{-k}}^{\dagger}+b_{1 ,\mathbf{-k}}  b_{2 ,\mathbf{k}}^{\dagger}\right)\\
    &\left.+g_{2}^{*}(\tau)\left(b_{2, \mathbf{-k}} b_{1, \mathbf{k}}^{\dagger}+b_{2,\mathbf{k}} b_{1, \mathbf{-k}}^{\dagger}\right)\right\}
    \end{aligned}
\end{equation}
Where, the terms $l_{1}$, $l_{2}$, $\omega_1$, $\omega_2$, $g_1$ and $g_2$ are defined below: 
\begin{equation}
  \begin{aligned}
  &l_{1}(\tau)=\frac{{K^{2}} \pi_{a}^{2}}{16 a^{4} m_{1}}+i \frac{3}{4 a^{2}} \pi_{a} \\
  &l_{2}(\tau)=\frac{{K^{2}} \pi_{a}^{2}}{16 a^{4} m_{2}}+i \frac{3}{4 a^{2}} \pi_{a} \\
  &\omega_{1}(\tau)=\frac{1}{2}\left(m_{1}+\frac{{K^{2}} \pi_{a}^{2}}{8 a^{4} m_{1}}\right) \\
  &\omega_{2}(\tau)=\frac{1}{2}\left(m_{2}+\frac{{K^{2}} \pi_{a}^{2}}{8 a^{4} m_{2}}\right) \\
  &g_{1}(\tau)=\frac{{K^{2}} \pi_{a}^{2}}{4 a^{4}} \frac{1}{\sqrt{m_{1} m_{2}}} \\
  &~~~~~~~~+i \frac{{K^{2}} \pi_{a}^{2}}{4 a^{4}} \frac{\left(m_{1}+m_{2}\right)}{\sqrt{m_{1} m}} =g_{2}(\tau)
  \end{aligned}
\end{equation}
Here it is important to note that,  $\tau$ represents the conformal time which is related to the physical time $t$ by the following expression:
\begin{eqnarray}
\tau= \int \frac{dt}{a(t)}=-\frac{1}{H}\exp(-Ht).
\end{eqnarray}
Consequently,  in terms of the conformal time coordinate the de Sitter metric in planar patch can be recast as:
\begin{eqnarray}
ds^{2}=a^{2}(\tau)\left(-d\tau^{2}+d{\bf x}^{2}\right)\quad {\rm where}\quad a(\tau)=-\frac{1}{H\tau}.\quad
\end{eqnarray}


\subsection{\textcolor{Sepia}{\textbf{The Two Coupled Inverted Harmonic Oscillator}}} \label{Inverte}

The two coupled inverted quantum harmonic oscillator \cite{Tarzi_1988,yuce2006inverted,SUBRAMANYAN2021168470} can be described by the Hamiltonian $\text{H}$ which is the sum of the free Hamiltonian of both the inverted quantum harmonic oscillators and the interaction Hamiltonian $\text{H}_{int}$ which depends on the type interaction between them and the coupling constant $K$ accounts for the strength of the interaction between the two coupled inverted quantum harmonic oscillators:

\begin{equation}\label{iqhc}
\text{H}=\sum_{i=1}^{2} \text{H}_{i} + K \text{H}_{int}
\end{equation}

Here, ${\text{H}}_{i}$ is the Hamiltonian of a free two inverted quantum harmonic oscillators:

\begin{equation}
\begin{aligned}
\text{H}_{i} &=\frac{\hat{p}_{i}^{2}}{2}-\frac{\hat{q}_{i}^{2}}{2}\quad {\rm where}\quad i=1,2\\
&=i \frac{\hbar}{2}\left(\hat{b}_{i}^{2} e^{2 i \frac{\pi}{4}}-\text { h.c. }\right) .
\end{aligned}
\end{equation}

Here $b_i$ corresponds to the annihilation operator for the quantum harmonic oscillator and index $i$ runs from 1 to 2.

 In the present context of discussion we will consider the construction of General squeeze Hamiltonian. Writing the free part Hamiltonian in this form facilitates the comparison with the more general squeeze Hamiltonian, which we will be considering. The setup of this two coupled inverted harmonic oscillator can be mapped to case of two coupled scalar fields in de Sitter background as indicated in the Eq\eqref{iqhc} and  Eq\eqref{hamilto}. The free part of the Hamiltonian for two coupled inverted quantum harmonic oscillator gets mapped to the terms containing $l_{1}$, $l^{*}_{1}$ for the first scalar field with modes $(1,{\bf{k}},1,{\bf{-k}}$) and the terms containing $l_{2}$, $l^{*}_{2}$ for the second scalar field with modes $(2,{\bf{k}},2,{\bf{-k}}$). The free terms with $\omega_{1}$ and $\omega_{2}$ mimics the role of rotation operator which is absent in Eq\eqref{iqhc} but it can be introduced explicitly in it. The interaction Hamiltonian $H_{int}$ corresponds to the terms containing $g_1$, $g^*_1$, $g_2$ and $g^*_2$.

\subsection{\textcolor{Sepia}{\textbf{Four mode Squeezed State operator}}} \label{fmss1}
Let us consider a state $\ket{\phi}_{\text{in}}$ which is the initial reference vacuum state of the two scalar fields. The final out state $\ket{\phi}_{\text{out}}$ can be obtained by by applying the operator given in Eq \eqref{eq10} on the initial reference vacuum state of the two scalar fields.
  
\begin{equation}
  \begin{aligned}
    \ket{\phi}_{\text {out }} =
    S_{1}^{(1)}\left(r_{1}, \theta_{1}\right) 
    S_{1}^{(2)}\left(r_{2}, \theta_{2}\right) 
    &S\left(r_{1},r_{2}, \theta_{1}, \theta_{2}\right) \\
    &\mathcal{R}\left(\phi_{1}, \phi_{2}\right)
    \ket{\phi}_{i n}
  \end{aligned}
  \label{eq10}
\end{equation}
Here, the initial vacuum state is $\ket{\phi}_{in}$ and we have
\begin{equation}
\mathcal{R}\left(\phi_{1}, \phi_{2}\right)|0,0\rangle =e^{i\left(\phi_{1}+\phi_{2}\right)}|0,0\rangle
\end{equation}
Here, we can see that the contribution of the total rotational operator on the vacuum state just introduces an overall phase factor and we neglect it for further calculations of the total squeezed operator for the two scalar fields in de Sitter background space.

We write the most general Squeezed state in the present context, which is defined as:
\begin{equation}
\begin{aligned}
\left|\Psi_{\text {sq }}\right\rangle=S_{1}^{(1)}\left(r_{1}, \theta_{1}\right) S_{1}^{(2)}\left(r_2, \theta_{2}\right)  S\left(r_{1},r_{2}, \theta_{1}, \theta_{2}\right)|0,0\rangle
\end{aligned}
\end{equation}

Here it is important to note that
\begin{equation}
  \begin{aligned}
    \underbrace{S_{1}^{(1)}\left(r_{1}, \theta_{1}\right) S_{1}^{(2)}\left(r_{2}, \theta_{2}\right)}_{\textcolor{blue}{\rm with~interaction~1~and~2}}& \neq
    &\underbrace{S_{1}^{(1)}\left(r_{1}, \theta_{1}\right) S_{1}^{(2)}\left(r_{2}, \theta_{2}\right)}_{\textcolor{blue}{\rm without~interaction~1~and~2}}
  \end{aligned}
\end{equation}
Following are the important points to be noted for operators $S_{1}^{(1)}\left(r_{1}, \theta_{1}\right) ,~ S_{1}^{(2)}\left(r_{2}, \theta_{2}\right)$ and $S\left(r_{1},r_{2}, \theta_{1}, \theta_{2}\right)$:
\begin{itemize}
  \item The contribution from the gravitational part from the $a(t)$ as well as the interaction part of the fields appear in $S\left(r_{1},r_{2}, \theta_{1}, \theta_{2}\right)$
  
  \item The Contribution of $\mu_{1}$ and $\mu_{2}$ with interaction appears in $S_{1}^{(1)}\left(r_{1}, \theta_{1}\right) ,~ S_{1}^{(2)}\left(r_{2}, \theta_{2}\right)$,  which makes them different from the contribution obtained from the case when there is no interaction at all.
\end{itemize}

\subsection{\textcolor{Sepia}{\textbf{ Technical details of the Constructions}}}
Here we will give the calculation of the total squeezed operator for two scalar fields having interaction with each other in de Sitter background space. First we consider the operator $S_{1}^{(1)}\left(r_{1}, \theta_{1}\right)$ and write it in the form of Eq\eqref{eq14} which is the product of usual squeezed operator for first field and the interaction term given in Eq\eqref{eq15}.
\begin{equation}
  \begin{aligned}
  &S_{1}^{(1)}\left(r_{1}, \theta_{1}\right) =\exp \left[\frac { r _ { 1 } } { 2 } \left(e^{-2 i \theta_{1}} b_{1}^{2}\right.\right.\left.-e^{2 i \theta_{1}} b_{1}^{\dagger 2}\right)\\
  &+\frac{r_{1}}{2}\left(e^{-2 i \theta_{1}}\left(b_{1} b_{2}+b_{1}^{\dagger} b_{2}^{\dagger}\right)\right.\left.-e^{2 i \theta_{1}}\left(b_{2}^{\dagger} b_{1}^{\dagger}+b_{2} b_{1}\right)\right)\\
  &=S_{1}^{\text{\tiny (non-int.)}}\left(r_{1}, \theta_{1}\right) \underbrace{S_{1}^{ \text{\tiny(1 int. with 2)}}\left(r_{1}, \theta_{1}\right)}_{\textcolor{blue}{\rm New~Contribution}}
  \end{aligned}
  \label{eq14}
  \end{equation}
where,  $S_{1}^{ \text{\tiny(1 int. with 2)}}\left(r_{1}, \theta_{1}\right)$ is the part of the operator $S_{1}^{(1)}\left(r_{1}, \theta_{1}\right)$ containing  interaction,  which is given by:
\begin{equation}
  \begin{aligned}
    S_{1}^{ \text{\tiny(1 int. with 2)}}\left(r_{1}, \theta_{1}\right) = &
    \exp \Big[ \frac {r_{1}}{2} \Big\{ e^{-2 i \theta_{1}}\left(b_{1} b_{2} 
  +b_{1}^{\dagger} b_{2}^{\dagger}\right) \\
  & -e^{2 i \theta_{1}}\left(b_{2}^{\dagger} b_{1}^{\dagger} + b_{2} b_{1}\right)\Big\} +\ldots ]
  \end{aligned}
  \label{eq15}
\end{equation}

Similarly we can write for operator $S_{1}^{(2)}\left(r_{2}, \theta_{2}\right)$, which is also the product of usual squeezed operator for second field and the term coming from the interaction. 
\begin{equation}
  \begin{aligned}
  S_{1}^{(2)}\left(r_{2}, \theta_{2}\right) &= \\
  &S_{1}^{\text{\tiny (non-int.)}}\left(r_{2}, \theta_{2}\right) \underbrace{S_{1}^{ \text{\tiny(2 int. with 1)}}\left(r_{2}, \theta_{2}\right)}_{\textcolor{blue}{\rm New~Contribution}}
  \end{aligned}
  \end{equation}
where, Eq\eqref{eq17} represents the interaction part of the operator $ S_{1}^{(2)}\left(r_{2}, \theta_{2}\right)$ coming from second field.
\begin{equation}
  \begin{aligned}
    S_{1}^{\text{\tiny (1 int. with 2)}}\left(r_{2}, \theta_{2}\right) = &
    \exp \Big[ \frac {r_{2}}{2} \Big\{ e^{-2 i \theta_{2}}\left(b_{1} b_{2} 
  +b_{1}^{\dagger} b_{2}^{\dagger}\right) \\
  & -e^{2 i \theta_{1}}\left(b_{2}^{\dagger} b_{1}^{\dagger} + b_{2} b_{1}\right)\Big\} +\ldots ]
  \end{aligned}
  \label{eq17}
\end{equation}

Now, the full squeezed state operator is the product of the usual squeezed operators from both the fields with the operators with interaction between 1 and 2 fields, a third additional term is also present denoted as $S\left(r_{1},r_{2}, \theta_{1}, \theta_{2}\right)$. This is  the contribution which is appearing due to commutators between $b_{1}$, $b_{1}^{\dagger}$, and $b_{2}$, $b_{2}^{\dagger}$. 

\begin{equation}
  \begin{aligned}
    S_{\text{Full}}=&\\
    &\underbrace{S_{1}\left(r_{1}, \theta_{1}\right)S_{1}\left(r_{2}, \theta_{2}\right)}_{\textcolor{blue}{\rm Without~interaction}}\\
    &\times ~~ \underbrace{S_{1}^{\text{int.}}\left(r_{1}, \theta_{1}\right)S_{1}^{\text{int.}}\left(r_{2}, \theta_{2}\right)}_{\textcolor{blue}{\rm Interaction~between~1~and~2}}~~\times\underbrace{S\left(r_{1}, \theta_{1}, r_{2}, \theta_{2}\right)}_{\textcolor{blue}{\rm Additional~terms}}
  \end{aligned}
  \label{eq18}
\end{equation}

In order to construct the additional term $S\left(r_{1},r_{2}, \theta_{1}, \theta_{2}\right)$ of the total squeezed operator we will use the Baker–Campbell–Hausdorff formula. Let us consider $ S_{1}^{(1)}$ and $ S_{2}^{(1)}$ given in Eq \eqref{eq19} and Eq \eqref{eq20}.

\begin{equation}
  \begin{aligned}
  S_{1}^{(1)}\left(r_{1}, \theta_{1}\right)=
  \exp &\bigg[\frac{r_{ 1 }}{ 2 } \left(e^{-2 i \theta_{1}} b_{1}^{2} - e^{2 i \theta_{1}} b_{1}^{\dagger_2}\right) \\
  + &\frac{r_{1}}{2} \bigg\{e^{-2 i \theta_{1}}\left(b_{1} b_{2}+b_{1}^{\dagger}b_{2}^{\dagger}\right)\\
  &~~~~~~~~~- e^{2 i \theta_{1}}\left(b_{2}^{\dagger} b_{1}^{\dagger}+b_{2} b_{1}\right)\bigg\}\bigg]
  \end{aligned}
  \label{eq19}
\end{equation}
and
\begin{equation}
  \begin{aligned}
  S_{1}^{(2)}\left(r_{1}, \theta_{2}\right)=
  \exp &\bigg[\frac{r_{ 2 }}{ 2 } \left(e^{-2 i \theta_{2}} b_{1}^{2} - e^{2 i \theta_{2}} b_{1}^{\dagger 2}\right) \\
  + &\frac{r_{2}}{2} \bigg\{e^{-2 i \theta_{2}}\left(b_{1} b_{2}+b_{1}^{\dagger}b_{2}^{\dagger}\right)\\
  &~~~~~~~~~- e^{2 i \theta_{2}}\left(b_{2}^{\dagger} b_{1}^{\dagger}+b_{2} b_{1}\right)\bigg\}\bigg]
  \end{aligned}
  \label{eq20}
\end{equation}

Now we call,
\begin{equation}
  \begin{aligned}
    &~~~~~~~~~~\frac{r_{ 1 }}{ 2 } \left(e^{-2 i \theta_{1}} b_{1}^{2} - e^{2 i \theta_{1}} b_{1}^{\dagger 2}\right) &= \alpha_{1}\\
    &\frac{r_{1}}{2} \bigg\{e^{-2 i \theta_{1}}\left(b_{1} b_{2}+b_{1}^{\dagger}b_{2}^{\dagger}\right)\\
   &~~~~~~~~~~~- e^{2 i \theta_{1}}\left(b_{2}^{\dagger} b_{1}^{\dagger}+b_{2} b_{1}\right)\bigg\}\bigg] &= \alpha_{2}
  \end{aligned}
\end{equation}
Similarly we have,
\begin{equation}
  \begin{aligned}
    &~~~~~~~~~~\frac{r_{ 2}}{ 2 } \left(e^{-2 i \theta_{2}} b_{1}^{2} - e^{2 i \theta_{2}} b_{2}^{\dagger 2}\right) &= \beta_{1}\\
    &\frac{r_{2}}{2} \bigg\{e^{-2 i \theta_{2}}\left(b_{1} b_{2}+b_{1}^{\dagger}b_{2}^{\dagger}\right)\\
   &~~~~~~~~~~~- e^{2 i \theta_{2}}\left(b_{2}^{\dagger} b_{1}^{\dagger}+b_{2} b_{1}\right)\bigg\}\bigg] &= \beta_{2}
  \end{aligned}
  \label{eq22}
\end{equation}

Such that 
\begin{equation}
  \begin{aligned}
    \alpha_{1}+\alpha_{2}=A_{1} \\
    \beta_{1}+\beta_{2}=A_{2}
  \end{aligned}
  \label{A1A2}
\end{equation}

Now we apply the Baker–Campbell–Hausdorff formula between $S_{1}^{(1)}$ and $ S_{2}^{(1)}$, where we have defined the terms in the exponential as $A_1$ and $A_2$.

\begin{equation}
  \begin{aligned}
    e^{A_{1}} \cdot e^{A_{2}}&=\\
    &e^{\{ A_{1} + A_{2} +\frac{1}{2} [A_{1},A_{2}] + \frac{1}{12}\left([A_{1},[A_{1},A_{2}]]-[A_{2},[A_{1},A_{2}]]\right) + \ldots \}}\\
    &=\underbrace{~e^{A_{1}}~}_{\textcolor{blue}{S_{1}^{1}}}~ \underbrace{~e^{A_{2}}~}_{\textcolor{blue}{S_{1}^{1}}}
    \underbrace{e^{f(A_{1},A_{2})}}_{\textcolor{blue}{S~ \rm New~Contribution}}
  \end{aligned}
  \label{eq24}
\end{equation}

We will only consider the BCH-expansion upto $\left[A_{1},\left[A_{1}, A_{2}\right]\right]$ and $\left[A_{2},\left[A_{1}, A_{2}\right]\right]$.
\begin{equation}
  \begin{aligned}
f\left(A_{1}, A_{2}\right)&= \\
  &\frac{1}{2}\left[A_{1}, A_{2}\right]+\frac{1}{12} \big(\left[A_{1},\left[A_{1}, A_{2}\right]\right]
  \\ 
  &~~~~~~~~~~~~~~~~~~~~~~-\left[A_{2},\left[A_{1}, A_{2}\right]\right]\big) \\
  &~~~~~~~~~~~~~~~~~~~~~~~~~~~+\cdots\\
\end{aligned}
\label{eq25}
\end{equation}

First we consider the first commutator in the Eq\eqref{eq25}
\begin{equation}
  \begin{aligned}
  \left[A_{1}, A_{2}\right]=&[\alpha_{1}+\alpha_{2},\beta_{1}+\beta_{2}]\\
  &\left[\alpha_{1}, \beta_{1}\right]+\left[\alpha_{2}, \beta_{2}\right] +\left[\alpha_{2}, \beta_{1}\right]+\left[\alpha_{1}, \beta_{2}\right]
  \end{aligned}
\end{equation}

Where,

\begin{equation}
  \begin{aligned}
  \left[\alpha_{1},\beta_{1}\right]&=0 \\
  \left[\alpha_{2},\beta_{2}\right]&=0 \\
  \left[\alpha_{2}, \beta_{1}\right]&=\frac{r_{1} r_{2}}{2}\left\{\left(e^{- 2 i\left(\theta_{1}+\theta_{2}\right)}\right.\right.\left.-e^{-2 i\left(\theta_{1}-\theta_{2}\right)}\right) b_{1} b_{2}^{\dagger}\\
  &~~~~~~~~~~~~\left.+\left(e^{2 i\left(\theta_{1}-\theta_{2}\right)}-e^{-2 i\left(\theta_{1}+\theta_{2}\right)}\right) b_{1}^{\dagger}b_{2}\right\}\\
  \left[\alpha_{1}, \beta_{2}\right]&=\frac{r_{1} r_{2}}{2}\left\{\left(e^{- 2 i\left(\theta_{1}+\theta_{2}\right)}\right.\right.\left.-e^{-2 i\left(\theta_{1}-\theta_{2}\right)}\right) b_{1} b_{2}^{\dagger}\\
  &~~~~~~~~~~~~\left.+\left(e^{2 i\left(\theta_{1}+\theta_{2}\right)}-e^{2 i\left(\theta_{1}-\theta_{2}\right)}\right) b_{1}^{\dagger}b_{2}\right\}
  \end{aligned}
\end{equation}

So, 
\begin{equation}
  \begin{aligned}
    \left[A_{1},A_{2}\right]&=[\alpha_{1},\beta_{2}] +[\alpha_{2},\beta_{1}]\\
    &=\frac{r_{1} r_{2}}{2}\left\{f_{1}\left(\theta_{1}, \theta_{2}\right)~b_{1} b_{2}^{\dagger}+f_{2}\left(\theta_{1}, \theta_{2}\right)~b_{1}^{\dagger} b_{2}\right\}
  \end{aligned}
 \end{equation}

 Where, we have defined $f_{1}\left(\theta_{1}, \theta_{2}\right)$ and $f_{2}\left(\theta_{1}, \theta_{2}\right)$ as:
 \begin{equation}
 \begin{aligned}
 f_{1}\left(\theta_{1}, \theta_{2}\right)&=2 \cos \left(2\left(\theta_{1}+\theta_{2}\right)\right) 
 -2 e^{-2 i\left(\theta_{1}-\theta_{2}\right)} \\
 &=f_{1}^{\operatorname{Real}}\left(\theta_{1}+\theta_{2}\right)+i f_{1}^{\rm I m}\left(\theta_{1}, \theta_{2}\right)
 \end{aligned}
\end{equation}

with the real and the imaginary part being,
\begin{equation}
  \begin{aligned}
    f_{1}^{\operatorname{Real}}\left(\theta_{1}, \theta_{2}\right)&=2\left(\cos \left(2\left(\theta_{1}+\theta_{2}\right)\right)-\cos(2(\theta_{1}-\theta_{2}))\right) \\
f_{1}^{\operatorname{Im}}\left(\theta_{1}, \theta_{2}\right)&=2 \sin \left(2\left(\theta_{1}-\theta_{2}\right)\right)
  \end{aligned}
\end{equation}

and 

\begin{equation}
  \begin{aligned}
    f_{2}\left(\theta_{1}, \theta_{2}\right) &=e^{2 i\left(\theta_{1}+\theta_{2}\right)}-e^{-2 i\left(\theta_{1}+\theta_{2}\right)} \\
    &= 2 i\sin\left(2\left(\theta_{1}+\theta_{2}\right)\right) \\
    &=f_{2}^{\rm Real}\left(\theta_{1}, \theta_{2}\right) +i f_{2}^{\rm Im}\left(\theta_{1}, \theta_{2}\right)
    \end{aligned}
\end{equation}

where,
\begin{equation}
  \begin{aligned}
    f_{2}^{\rm Real} \left(\theta_{1}, \theta_{2}\right)&=0 \\
    f_{2}^{I m}\left(\theta_{1}, \theta_{2}\right)&=2 \sin \left(2\left(\theta_{1}+\theta_{2}\right)\right)
  \end{aligned}
\end{equation}

So, the first commutator becomes as follows:
\begin{equation}
  \begin{aligned}
  \left[A_{1}, A_{2}\right]&=r_{1} r_{2} \left[\left\{ \cos \left(2\left(\theta_{1}+\theta_{2}\right)\right)-\cos(2(\theta_{1}-\theta_{2})) \right.\right.\\
  &~~~~~~~~~~~~\left.+i\sin(2(\theta_{1}+\theta_{2}))
  \right\} b_{1} b_{2}^{\dagger}\\
  &~~~~~~~~~~~~~~~~~\left.+\left\{i \sin(2(\theta_{1}-\theta_{2})) \right\} b_{1}^{\dagger}b_{2}\right] \\
  \end{aligned}  
\end{equation}

We can express it as follows:
\begin{equation}
  \begin{aligned}
    \left[A_{1}, A_{2}\right]=&\left[f_{1}\left(r_{1}, r_{2}, \theta_{1}, \theta_{2}\right) b_{1}b_{2}^{\dagger}\right.\\
    &~~~~~~~~~~~~~~~~\left.+f_{2}\left(r_{1}, r_{2}, \theta_{1}, \theta_{2}\right) b_{1}^{\dagger} b_{2}\right]
    \end{aligned}
\end{equation}
Where, we define $f_{1}\left(r_{1}, r_{2}, \theta_{1}, \theta_{2}\right)$ and $f_{2}\left(r_{1}, r_{2}, \theta_{1}, \theta_{2}\right)$ as:
\begin{equation}
  \begin{aligned}
  f_{1}\left(r_{1}, r_{2}, \theta_{1}, \theta_{2}\right) &= r_{1}r_{2}\left[\left\{ \cos \left(2\left(\theta_{1}+\theta_{2}\right)\right)-\cos(2(\theta_{1}-\theta_{2}))\right. \right.  \\
  &
 +i \sin(2(\theta_{1}-\theta_{2}))]
\\
  f_{2}\left(r_{1}, r_{2}, \theta_{1}, \theta_{2}\right)&= r_{1} r_{2} i[ \sin(2(\theta_{1}+\theta_{2}))]   \\  \end{aligned}  
\end{equation}

Let us denote the following,

\begin{equation}
  \begin{aligned}
    P= f_{1}\left(r_{1}, r_{2}, \theta_{1}, \theta_{2}\right)b_{1}b_{2}^{\dagger}\\
   Q=f_{2}\left(r_{1}, r_{2}, \theta_{1}, \theta_{2}\right)b_{1}^{\dagger}b_{2}
  \end{aligned}
\end{equation}

So, $[A_{1},A_{2}]=P+Q$. We will now compute the second commutator in the expansion which is $\left[A_{1},\left[A_{1}, A_{2}\right]\right]$. The first part of this commutator is $\left[\alpha_{1},\left[A_{1},A_{2}\right]\right]$, which becomes:

\begin{equation}
  \begin{aligned}
    \left[\alpha_{1},\left[A_{1},A_{2}\right]\right]\\
    &= \left[\alpha_{1},P + Q \right]\\
    &\left(\tilde{f_{2}}b_{1}b_{2}+\tilde{f_{1}}b_{1}^{\dagger}b_{2}^{\dagger}\right)
\end{aligned}
\end{equation}
Where
\begin{equation}
  \begin{aligned}
    &\tilde{f_{2}} = r_{1}f_{2} e^{-2 i \theta_{1}}\\
    &\tilde{f_{1}}=r_{1}f_{1}e^{2i\theta_{1}}
  \end{aligned}
\end{equation}

Now we will compute the second part of the commutator $\left[A_{1},\left[A_{1}, A_{2}\right]\right]$ which is given as $[\alpha_{2},[A_{1},A_{2}]]$,
Here, first we write $\alpha_{2}$ as,

\begin{equation}
  \alpha_{2} = \Theta_{1}-\Theta_{2}
\end{equation}
Where,

\begin{equation}
  \begin{aligned}
    \Theta_{1} &=\frac{r_{1}}{2} e^{-2 i \theta_{1}}\left(b_{1} b_{2}+b_{1}^{\dagger}b_{2}^{\dagger}\right)\\
    \Theta_{2} &=\frac{r_{1}}{2}e^{2 i \theta_{1}}\left(b_{2}^{\dagger} b_{1}^{\dagger}+b_{2} b_{1}\right)
  \end{aligned}
\end{equation}

Using this we have,
\begin{equation}
  \begin{aligned}
    \left[\alpha_{2},\left[A_{1}, A_{2}\right]\right]&\\
&=\left[\Theta_{1}-\Theta_{2}, P+Q\right]\\
  \end{aligned}
\end{equation}

where the commutators are as follows:
\begin{equation}
\begin{aligned}
  \left[\Theta_{1}, P\right]&=\frac{r_{1} f_{1}}{2} e^{-2 i \theta_{1}}\left(b_{1} b_{1}-b_{2}^{\dagger} b_{2}^{\dagger}\right)\\
 \left[\Theta_{1}, Q\right]&=\frac{r_{1} f_{2}}{2} e^{-2 i \theta_{1}}\left(b_{2} b_{2}-b_{1}^{\dagger} b_{1}^{\dagger}\right)\\
 \left[\Theta_{2}, P\right]&=\frac{r_{1} f_{1}}{2} e^{2 i \theta_{1}}\left(b_{1} b_{1}-b_{2}^{\dagger} b_{2}^{\dagger}\right)\\
 \left[\Theta_{2}, Q\right]&=\frac{r_{1} f_{2}}{2} e^{2 i \theta_{1}}\left(b_{2} b_{2}-b_{1}^{\dagger}b_{1}^{\dagger}\right)\\
\end{aligned}
\end{equation}
The second part of the commutator $\left[A_{1},\left[A_{1}, A_{2}\right]\right]$ becomes:
\begin{equation}
\begin{aligned}
  &\left[\alpha_{2}\left[A_{1},A_{2}\right]\right]=\\
  &\frac{r_{1}}{2}\left[e^{2 i \theta_{1}}\left\{f_{2}\left(b_{1}^{\dagger}b_{1}^{\dagger}-b_{2} b_{2}\right)
  -f_{1}\left(b_{1} b_{1}-b_{2}^{\dagger} b_{2}^{\dagger}\right)\right\}\right. \\
  &~~~~~~~~~~~~\left.+e^{-2 i \theta_{1}}\left\{f_{1}\left(b_{1} b_{1}-b_{2}^{\dagger}b_{2}^{\dagger}\right)
  +f_{2}\left(b_{2} b_{2}-b_{1}^{\dagger} b_{1}^{\dagger}\right)\right\}\right]
  \end{aligned}
\end{equation}

For the third and the last commutator of our truncated BCH-expansion, we need to calculate $\left[A_{2},\left[A_{1}, A_{2}\right]\right]$, and previously we had defined,  $A_{2}=\beta_{1}+\beta_{2}$ \eqref{A1A2}
\newline
Where, we define $\beta_1$ and $\beta_2$ as:

\begin{equation}
  \begin{gathered}
    \beta_{1}=M_{1}-M_{2} \\
\beta_{2}=N_{1}-N_{2} \\
  \end{gathered}
  \label{eq44}
\end{equation}
and hence, from Eq\eqref{eq22} and \eqref{eq44},
$M_1$ and $M_2$ becomes:
\begin{equation}
\begin{gathered}
M_{1}=\frac{r_{2}}{2} e^{-2 i \theta_{2}} b_{2}^{2}, \quad N_{1}=\frac{r_{2}}{2}\left(e^{-2 i \theta_{2}}\left(b_{1} b_{2}+b_{1}^{\dagger} b_{2}^{\dagger}\right)\right) \\
M_{2}=\frac{r_{2}}{2} e^{2 i \theta_{2}} b_{2}^{\dagger 2}, \quad N_{2}=\frac{r_{2}}{2}\left(e^{2 i \theta_{2}}\left(b_{2}^{\dagger} b_{1}^{\dagger}+b_{2} b_{1}\right)\right)
\end{gathered}
\end{equation}
With the following commutation relation in Eq\eqref{eq46} , we can write the first part of the $\left[A_{2},\left[A_{1}, A_{2}\right]\right]$ commutator.

\begin{equation}
\begin{gathered}
  \left[M_{2}, P\right]=0 \quad
  \left[M_{1}, P\right]=r_{2} f_{1} e^{-2 i \theta_{2}}  b_{1} b_{2} \\
  \left[M_{1}, Q\right]=0 \quad
  \left[M_{2}, Q\right]=-r_{2} f_{2} e^{2 i \theta_{2}} b_{1}^{\dagger} b_{2}^{\dagger} 
\end{gathered}
\label{eq46}
\end{equation}
Which is given as follows:
\begin{equation}
  \begin{aligned}
\left[\beta_{1},\left[A_{1}, A_{2}\right]\right]&=\left[\beta_{1}, P\right]+\left[\beta_{1}, Q\right] \\
  &=r_{2}\left(f_{1} e^{-2 i \theta_{2}} b_{1} b_{2}+f_{2} e^{2 i \theta_{2}} b_{1}^{\dagger} b_{2}^{\dagger}\right)
  \end{aligned}
\end{equation}

Similarly we calculate the second term of the commutator $\left[A_{2},\left[A_{1}, A_{2}\right]\right]$ and use the following commutation results given in the Eq\eqref{eq48} below,
\begin{equation}
  \begin{aligned}
    &\left[N_{1}, P\right]=\frac{r_{2} f_{1}}{2} e^{-2 i \theta_{2}}\left(b_{1} b_{1}-b_{2}^{\dagger} b_{2}^{\dagger}\right)\\
    &\left[N_{1}, Q\right]=\frac{r_{2} f_{2}}{2} e^{-2 i \theta_{2}}\left(b_{2} b_{2}-b_{1}^{\dagger} b_{1}^{\dagger}\right)\\
    &\left[N_{2}, P\right]=\frac{r_{2} f_{1}}{2} e^{2 i \theta_{2}}\left(b_{1} b_{1}-b_{2}^{\dagger} b_{2}^{\dagger}\right)\\
    &\left[N_{2}, Q\right]=\frac{r_{2} f_{2}}{2} e^{2 i \theta_{2}}\left(b_{2} b_{2}-b_{1}^{\dagger} b_{1}^{\dagger}\right)\\
    \end{aligned}
    \label{eq48}
\end{equation}
Finally we get the second term of the $\left[A_{2},\left[A_{1}, A_{2}\right]\right]$ commutator:
\begin{equation}
  \begin{aligned}
    &\left[\beta_{2},\left[A_{1}, A_{2}\right]\right]=\left[\beta_{2}, P+Q\right]\\
    &=\left[N_{1}, P\right]-\left[N_{2}, P\right]+\left[N_{1}, Q\right]-\left[N_{2}, P\right]\\
    &= i~r_{2} \sin \left(2 \theta_{2}\right)\left[f_{2}\left(b_{1}^{\dagger}b_{1}^{\dagger}-b_{2}b_{2}\right)\right.\\
    &~~~~~~~~~~~~~~~~~~~~~~~~~~~~~\left.-f_{1}\left(b_{1} b_{1}-b_{2}^{\dagger}b_{2}^{\dagger}\right)\right]
  \end{aligned}
\end{equation}

After calculating the terms in Eq\eqref{eq25}, now we rewrite the combination of $\left[A_{1},\left[A_{1}, A_{2}\right]\right]$ and $\left[A_{2},\left[A_{1}, A_{2}\right]\right]$ commutators in terms of $X_1$, $X_2$, $X_3$ and $X_4$.
\begin{equation}
\begin{aligned}
  &{\left[A_{1},\left[A_{1}, A_{2}\right]\right]-\left[A_{2}\left[A_{1}, A_{2}\right]\right]} \\
  &=X_{1} b_{1} b_{2}+X_{2} b_{1}^{\dagger} b_{2}^{\dagger} +X_{3}\left(b_{1}^{\dagger} b_{1}^{\dagger}-b_{2} b_{2}\right) \\
  &~~~+X_{4}\left(b_{1} b_{1}-b_{2}^{\dagger} b_{2}^{\dagger}\right) \\
\end{aligned}
\end{equation}

Where, $X_1$, $X_2$, $X_3$ and $X_4$ are appended below:
\begin{equation}
\begin{aligned}
&X_{1}=\left(r_{1} f_{2} e^{-2 i \theta_{1}}-r_{2} f_{1} e^{-2 i \theta_{2}}\right) \\
  &X_{2}=\left(r_{1} f_{1} e^{2 i \theta_{1}}-r_{2} f_{2} e^{2 i \theta_{2}}\right) \\
  &X_{3}=i f_{2}\left(r_{1} \sin \left(2 \theta_{1}\right)-r_{2} \sin \left(2 \theta_{2}\right)\right) \\
  &X_{4}=i f_{1}\left(r_{2} \sin \left(2 \theta_{2}\right)-r_{1} \sin \left(2 \theta_{1}\right)\right) \\
\end{aligned}
\end{equation}
  
Note: Here the ratio of $X_3$ and $X_4$ satisfies the following relation given in Eq\eqref{eq52} :
\begin{equation}
  \frac{X_{3}}{X_{4}}=-\frac{f_{2}}{f_{1}} 
  \label{eq52}
\end{equation}

To get the full squeezed operator for our case whose form is mentioned in Eq\eqref{eq18}, we multiply the usual free squeezed operators from both the fields with the the BCH expansion given in Eq\eqref{eq24}. The total squeezed operator for two scalar fields in de Sitter background space after applying the Baker-Campbell-Hausdorff will become:
\begin{equation}
  S_{\text{Full}}\left(r_{1}, \theta_{1}, r_{2}, \theta_{2}\right)=e^{\left(M_{1}+M_{2}+M_{3}\right)}  
\end{equation}
Where, $\exp M_1$, and $\exp M_2$ are the usual free squeezed operators given as:
\begin{equation}
  \begin{aligned}
\exp M_{1}&=\underbrace{\exp \left[\frac{r_{1}}{2}\left(e^{-2 i \theta_{1}} b_{1}^{2}-e^{2 i \theta_{1}} b_{1}^{\dagger 2}\right)\right]}_{\textcolor{blue}{ _ { for ``1" . }}}\\
\exp M_{2}&=\underbrace{\exp \left[\frac{r_{2}}{2}\left(e^{-2 i \theta_{2}} b_{2}^{2}-e^{2 i \theta_{2}} b_{2}^{\dagger 2}\right)\right]}_{\textcolor{blue}{ _ { for ``2" . }}}\\
\end{aligned}
\end{equation}
and $\exp M_3$ is the operator term which is coming due to the interaction between both the fields but we will introduce the coupling constant $K$ as an overall factor in the exponential which will help us in doing our further analysis perturbatively in the limit $K<< 1$.
\begin{equation}
  \begin{aligned}
\exp {KM_{3}}&=\exp {K(B_{1}+B_{2}+B_{3}+B_{4}+B_{5}+B_{6})}
  \end{aligned}
\end{equation}
We will introduce the coupling constant $K$ as an overall factor in the exponential in the above equation Where the various $B$ terms in the exponential of $M_3$ are:
\begin{equation}
  \begin{aligned}
    &B_{1}=\frac { r _ { 1 } } { 2 } \left[e^{-2 i \theta_{1}}\left(b_{1} b_{2}+b_{1}^{\dagger} b_{2}^{\dagger}\right) -e^{2 i \theta_{1}}\left(b_{2}^{\dagger} b_{1}^{\dagger}+b_{2} b_{1}\right)\right] \\
    &B_{2}=\frac{r_{2}}{2}\left[e^{-2 i \theta_{2}}\left(b_{1} b_{2}+b_{1}^{\dagger} b_{2}^{\dagger}\right)\right.\left.-e^{2 i \theta_{2}}\left(b_{2}^{\dagger} b_{1}^{\dagger}+b_{2} b_{1}\right)\right] \\
    &B_{3}= [f_{1}\left(r_{1}, r_{2}, \theta_{1}, \theta_{2}\right) b_{1} b_{2}^{\dagger} \left.+f_{2}\left(r_{1}, r_{2}, \theta_{1}, \theta_{2}\right) b_{1}^{\dagger} b_{2}\right]\\
    &B_{4}=\frac{1}{12}\left[\left(X_{1}\left(r_{1}, r_{2}, \theta_{1}, \theta_{2}\right) b_{1} b_{2}\right.+X_{2}\left(r_{1}, r_{2}, \theta_{1}, \theta_{2}\right) b_{1}^{\dagger} b_{2}^{\dagger}\right]\\
    &B_{5}=X_{3}\left(r_{1}, r_{2}, \theta_{1}, \theta_{2}\right)\left(b_{1}^{\dagger} b_{1}^{\dagger}-b_{2} b_{2}\right) \\
    &B_{6}=X_{4}\left(x_{1}, r_{2}, \theta_{1}, \theta_{2}\right)\left(b_{1} b_{1}-b_{2}^{\dagger} b_{2}^{\dagger}\right)\\
    \end{aligned}
\end{equation}

From the next section we will change the notations for squeezing parameters and replace $r_1\xrightarrow{}R_1$, $\phi_1\xrightarrow{}\Phi_1$, $\theta_1\xrightarrow{}\Theta_1$ and similarly for $r_2,~\phi_2, \theta_2 $ parameters.



\section{\textcolor{Sepia}{\textbf{ \Large Calculation for unitary evolution}}}\label{LCDE}

To understand the dynamics of the two scalar fields with interaction in de Sitter background space, we construct the most generic evolution operator $\mathcal{U}(\eta_{1},\eta_{2})$ which is the product of total squeezed operator and the total rotational operator for both the fields. The total rotational operator is defined as follows:
\begin{equation}
\begin{aligned}
\mathcal{R}_{\text{Total}}(\Phi_{1}, \Phi_{2})= \mathcal{R}_{1}(\Phi_{1})  \mathcal{R}_{2}(\Phi_{2})
    \end{aligned}
\end{equation}

\begin{equation}\label{unievo}
\begin{aligned}
    \mathcal{U}(\eta_{1},\eta_{2})&=\\
    &S_{1}\left(R_{1}, \Theta_{1}\right) S_{2}\left(R_{2}, \Theta_{2}\right) S_{12}\left(R_{1},R_{2}, \Theta_{1}, \Theta_{2}\right)\times\\
&~~~~~~~~~~~~~~~~~~~~~~~~~~~~~~~~\mathcal{R}_{1}(\Phi_{1})  \mathcal{R}_{2}(\Phi_{2})
    \end{aligned}
\end{equation}
In the Heisenberg picture representation the operators can be $\hat{v}_{1}(\mathbf{x}, \eta)$,  $\hat{\pi}_{1}(\mathbf{x}, \eta)$, $\hat{v}_{2}(\mathbf{x}, \eta)$ and $\hat{\pi}_{2}(\mathbf{x}, \eta)$ can be written as follows

\begin{equation}
\begin{aligned}
&\hat{v}_{1}(\mathbf{x}, \eta)=\mathcal{U}^{\dagger}\left(\eta, \eta_{0}\right) \hat{v}_{1}\left(\mathbf{x}, \eta_{0}\right) \mathcal{U}\left(\eta, \eta_{0}\right)\\
&=\int \frac{d^{3} k}{(2 \pi)^{3 / 2}} e^{i \mathbf{k} \cdot \mathbf{x}}\left(u_{1,\mathbf{k}}^{*}\left(\eta\right) b_{1,\mathbf{k}}+u_{1, -\mathbf{k}}\left(\eta\right) b_{1,-\mathbf{k}}^{\dagger}\right) \\
&\hat{\pi}_{1}(\mathbf{x}, \eta)=\mathcal{U}^{\dagger}\left(\eta, \eta_{0}\right) \hat{\pi}_{1}\left(\mathbf{x}, \eta_{0}\right) \mathcal{U}\left(\eta, \eta_{0}\right)\\
&=\int \frac{d^{3} k}{(2 \pi)^{3 / 2}} e^{i \mathbf{k} \cdot \mathbf{x}}\left(w_{1,\mathbf{k}}^{*}\left(\eta\right) b_{1,\mathbf{k}}+w_{1, -\mathbf{k}}\left(\eta\right) b_{1,-\mathbf{k}}^{\dagger}\right)\\
&\hat{v}_{2}(\mathbf{x}, \eta)=\mathcal{U}^{\dagger}\left(\eta, \eta_{0}\right) \hat{v}_{2}\left(\mathbf{x}, \eta_{0}\right) \mathcal{U}\left(\eta, \eta_{0}\right)\\
&=\int \frac{d^{3} k}{(2 \pi)^{3 / 2}} e^{i \mathbf{k} \cdot \mathbf{x}}\left(u_{2,\mathbf{k}}^{*}\left(\eta\right) b_{2,\mathbf{k}}+u_{2, -\mathbf{k}}\left(\eta\right) b_{2,-\mathbf{k}}^{\dagger}\right) \\
&\hat{\pi}_{2}(\mathbf{x}, \eta)=\mathcal{U}^{\dagger}\left(\eta, \eta_{0}\right) \hat{\pi}_{2}\left(\mathbf{x}, \eta_{0}\right) \mathcal{U}\left(\eta, \eta_{0}\right)\\
&=\int \frac{d^{3} k}{(2 \pi)^{3 / 2}} e^{i \mathbf{k} \cdot \mathbf{x}}\left(w_{2,\mathbf{k}}^{*}\left(\eta\right) b_{2,\mathbf{k}}+w_{2, -\mathbf{k}}\left(\eta\right) b_{2,-\mathbf{k}}^{\dagger}\right)
\end{aligned}
\label{mo1}
\end{equation}
In Schr\"odinger representation the position and momentum operators $v_{1,\textbf{k}}$, $\pi_{1,\textbf{k}}$ and $v_{2,\textbf{k}}$, $\pi_{2,\textbf{k}}$ for both the scalar fields in terms of annihilation and creation operator can be written in the following manner:

\begin{equation}
\begin{aligned}
&v_{1,\textbf{k}}\left(\eta\right)=\frac{1}{\sqrt{2 k}_{1}}\left(b_{1,\textbf{k}}\left(\eta\right)+b_{1,-\textbf{k}}^{\dagger}\left(\eta\right)\right)\\
&\pi_{1,\textbf{k}}\left(\eta\right)=i \sqrt{\frac{k_{1}}{2}}\left(b_{1,\textbf{k}}\left(\eta\right)-b_{1,-\textbf{k}}^{\dagger}\left(\eta\right)\right) \\
&v_{2,\textbf{k}}\left(\eta\right)=\frac{1}{\sqrt{2 k}_{2}}\left(b_{2,\textbf{k}}\left(\eta\right)+b_{2,-\textbf{k}}^{\dagger}\left(\eta\right)\right)\\
&\pi_{2,\textbf{k}}\left(\eta\right)=i \sqrt{\frac{k_{2}}{2}}\left(b_{2,\textbf{k}}\left(\eta\right)-b_{2,-\textbf{k}}^{\dagger}\left(\eta\right)\right)
\end{aligned}
\label{60eq}
\end{equation}
Where $b_{1,\textbf{k}}\left(\eta\right)$, $b_{1,-\textbf{k}}(\eta)$, $b_{2,\textbf{k}}(\eta)$ and $b_{2,-\textbf{k}}(\eta)$ are the annihilation and creation operators in time dependent Heisenberg representation for the two coupled scalar fields in de Sitter background space and using the factorized representation of the unitary time evolution operator introduced in the Eq\eqref{unievo}, the expression for the annihilation operator for field 1 can be written at any arbitrary time scale as:

\begin{widetext}
\begin{equation}\label{eq63e}
    \begin{aligned}
    b_{1}(\eta) &\equiv \mathcal{U}^{\dagger}\left(\eta, \eta_{0}\right) b_{2} \mathcal{U}\left(\eta, \eta_{0}\right)\\
    &={\mathcal{R}}^{\dagger}_{1}(\Phi_{1})  \mathcal{R}^{\dagger}_{2}(\Phi_{2}) S_{12}^{\dagger}\left(R_{1},R_{2}, \Theta_{1}, \Theta_{2}\right)
    S_{2}^{\dagger}\left(R_{2}, \Theta_{2}\right)
    S_{1}^{\dagger}\left(R_{1}, \Theta_{1}\right)
    b_{1} S_{1}\left(R_{1}, \Theta_{1}\right)
    S_{2}\left(R_{2}, \Theta_{2}\right)
    S_{12}\left(R_{1},R_{2}, \Theta_{1}, \Theta_{2}\right)\\
    &\times \mathcal{R}_{1}(\Phi_{1})\mathcal{R}_{2}(\Phi_{2})\\
    &={\mathcal{R}}^{\dagger}_{1}(\Phi_{1})  \mathcal{R}^{\dagger}_{2}(\Phi_{2})(\left(\cosh R_{1}\right) S_{12}^{\dagger} b_{1} S_{12}-\left(\sinh R_{1}\right) e^{2 \Theta_{1}} S_{12}^{\dagger} b_{1}^{\dagger} S_{12}){\mathcal{R}}_{1}(\Phi_{1})  \mathcal{R}_{2}(\Phi_{2})
\end{aligned}
\end{equation}
\end{widetext}

Using the Baker–Campbell–Hausdorff formula upto linear order in coupling constant $K$, the annihilation operator $b_{1}$ becomes \eqref{b1evo}. The total expression is the sum of the free part which is the usual time dependent expression for the annihilation operator on applying the squeezed operator (without interaction term) and there are extra terms with factor $K$, coming due to the interaction between both the fields. For the creation operator $b^{\dagger}_1$ of field 1, one can take the conjugate of Eq\eqref{eq63e} or \eqref{b1evo}.

\begin{widetext}
\begin{equation}\label{b1evo}
  \begin{aligned}
b_{1}(\eta)  &=\left(\cosh R_{1}e^{-i\Phi_{1}} b_{1}-e^{i(\Phi_1+2 \Theta_{1})} \sinh R_{1} b_{1}^{\dagger}\right)+K\bigg\{\Big(-\cosh R_{1}\left(i \left(R_{1} \sin 2 \Theta_{1} +\right.\right.\left.\left.+R_{2} \sin 2 \Theta_{2}\right)\right)-\frac{X_{2}}{12}\\
&-e^{2 i \Theta_{1}}\left(\sinh R_{1}\right)\left(f_{1}\right)\Big) e^{i\Phi_{2}} b_{2}^{\dagger}+ \left[\left(\cosh R_{1}\right)f_{2}+e^{2 i \Theta_{1}}\left(i\left(R_{1} \sin 2 \Theta_{1}+R_{2} \sin 2 \Theta_{2}\right)\right)-\frac{X_{1}}{12}\right](\sinh R_{1})e^{-i\Phi_{2}} b_{2}\\
&+\left[\frac{\cosh R_{1}}{2}\left(R_{1} \sin 2 \Theta_{1}+R_{2} \sin 2 \Theta_{2}\right)^{2}-\sinh R_{1} e^{2i \Theta_{1}}\right. \left.\left(\frac{X_{4}}{6}\right)\right] e^{-i\Phi_{1}}b_{1}+\left[\cosh R_{1}\left(\frac{X_{3}}{6}\right)-\frac{e^{2 i \Theta_{1}}}{2} 
  \sinh R_{1} \times\right.\\&\left.(R_{1} \sin 2\Theta_{1}\right. \left.\left.+R_{2} \sin 2 \Theta_{2}\right)^{2}\right] e^{i\Phi_{1}}b_{1}^{\dagger}\bigg\}+\ldots 
  \end{aligned}
  \end{equation}
\end{widetext}

Now for operator $b_2$, we will apply the same factorized representation of the unitary time evolution operator mentioned in the Eq\eqref{unievo}, the expression for the annihilation operator for field 2 can be written at any arbitrary time scale as Eq\eqref{65eqe}. For the creation operator $b^{\dagger}_2$ of field 2, one can take the conjugate of Eq\eqref{65eqe} or \eqref{b2evo}.

\begin{widetext}
\begin{equation}\label{65eqe}
    \begin{aligned}
    b_{2}(\eta) &\equiv \mathcal{U}^{\dagger}\left(\eta, \eta_{0}\right) b_{2} \mathcal{U}\left(\eta, \eta_{0}\right)\\
    &={\mathcal{R}}^{\dagger}_{1}(\Phi_{1})  \mathcal{R}^{\dagger}_{2}(\Phi_{2})S_{12}^{\dagger}\left(R_{1},R_{2}, \Theta_{1}, \Theta_{2}\right)
    S_{2}^{\dagger}\left(R_{2}, \Theta_{2}\right)
    S_{1}^{\dagger}\left(R_{1}, \Theta_{1}\right)
    b_{2} S_{1}\left(R_{1}, \Theta_{1}\right)
    S_{2}\left(R_{2}, \Theta_{2}\right)
    S_{12}\left(R_{1},R_{2}, \Theta_{1}, \Theta_{2}\right)\times\\&{\mathcal{R}}_{1}(\Phi_{1})  \mathcal{R}_{2}(\Phi_{2})={\mathcal{R}}^{\dagger}_{1}(\Phi_{1})  \mathcal{R}^{\dagger}_{2}(\Phi_{2})(\left(\cosh R_{1}\right) S_{12}^{\dagger} b_{2} S_{12}-\left(\sinh R_{1}\right) e^{2 \Theta_{1}} S_{12}^{\dagger} b_{2}^{\dagger} S_{12}){\mathcal{R}}_{1}(\Phi_{1})  \mathcal{R}_{2}(\Phi_{2})
\end{aligned}
\end{equation}
\end{widetext}

Using the same Baker–Campbell–Hausdorff formula upto linear order in coupling constant $K$, for the annihilation operator $b_{2}$ becomes \eqref{b2evo}. Here also we see that the total expression is sum of the free part which is the usual time dependent expression for the annihilation operator on applying the squeezed operator and the terms with factor $K$, coming due the presence of $S\left(R_{1}, \Theta_{1}, R_{2}, \Theta_{2}\right)$ operator which is accounting for the interaction between the two fields.

\begin{widetext}
\begin{equation}\label{b2evo}
  \begin{aligned}
   b_{2}(\eta) &=\left(\cosh R_{2}e^{-i\Phi_{2}} b_{2}-e^{i(\Phi{2}+2\Theta_{2})} \sinh R_{2} b_{2}^{\dagger}\right)+K\Big(\Big\{-i \cosh R_{2}\left(R_{1} \sin 2 \Theta_{1} +\right. \left.+R_{2} \sin 2 \Theta_{2}\right)-\cosh R_{2}\left(\frac{X_{2}}{12}\right)\\&-e^{2 i \Theta_{2}} \sinh R_{2}\left(f_{2}\right)\Big\} e^{i\Phi_{1}} b_{1}^{\dagger}+\Big(f_{1} \cosh R_{2}+e^{2 i \Theta_{2}} \sinh R_{2}\left(-i\left(R_{1} \sin 2 \Theta_{1}+R_{2} \sin 2 \Theta_{2}\right)\right. \left.+\frac{X_{1}}{12}\right) e^{-i\Phi_{1}} b_{1}+ \\
    &\left\{\left(R_{1} \sin 2 \Theta_{1}+R_{2} \sin 2 \Theta_{2}\right)^{2} \frac{\cosh R_{2}}{2}+\right.\left.e^{2 i \Theta_{2}} \sinh R_{2}\left(\frac{X_{3}}{6}\right)\right\} e^{-i\Phi_{2}} b_{2}+\left\{-\left(R_{1} \sin 2 \Theta_{1}+R_{2} \sin 2 \Theta_2\right)^{2}\right.\left.\right.\sinh R_{2}\times \\
    &\left.\left(\frac{e^{2 i \Theta_{2}}}{2}\right)-\frac{X_{4}}{6} \cosh R_{2}\right\} e^{i\Phi_{2}} b_{2}^{\dagger}\Big)+\ldots
    \end{aligned}  
\end{equation}
\end{widetext}

\section{\textcolor{Sepia}{\textbf{ \Large Evolution equations}}} \label{eqdiff1}
After getting the time evolution of operators $b_1$ and $b_2$, we compute the expressions for the differential equation of mode functions $u_{1,\bf{k}}$, $u_{2,\bf{k}}$,  $w_{1,\bf{k}}$ and  $w_{2,\bf{k}}$ for the two coupled scalar fields in de Sitter background space because it is crucial step for getting the evolution equation in terms of differential equation for squeeze factor $R_{1,\bf{k}},R_{2,\bf{k}}$, squeeze phase $\Theta_{1,\bf{k}},\Theta_{2,\bf{k}}$ and squeeze angle $\Phi_{1,\bf{k}},\Phi_{2,\bf{k}}$.
For calculating the differential equation we apply the Heisenberg equation of motion  Eq\eqref{heis} for different position and momentum operators. Using the Heisenberg equation of motion we will find that the mode functions $u_{1,\bf{k}}$, $u_{2,\bf{k}}$,  $w_{1,\bf{k}}$ and  $w_{2,\bf{k}}$ satisfy the Hamilton equations Eq\eqref{modediff}
This set of four differential equation are the classical equation of motion. There are 2 pairs one for each scalar fields the two pairs are symmetric with each other (only differ in indexing variable) but they are also coupled differential equation equation with coupling constant $K$. These equations tells us the dynamics of the position and momentum variables of the classical two interacting scalar field theory given by the action in Eq\eqref{acti}.

\begin{equation} \label{heis}
    \frac{d}{d t} \mathcal{O}_{\mathrm{i}}=i\left[\mathcal{H}, \mathcal{O}_{\mathrm{i}}\right]
\end{equation}
Where
$\begin{aligned}
\mathcal{O}_{\mathrm{i}} = \{v_{1}, \pi_{1}, v_{2}, \pi_{2}\}.
\end{aligned}$

\begin{equation}
\begin{aligned}
u_{1,\bf{k}}^{\prime}=&~~ w_{1,\bf{k}}-\left(\frac{P_{a}^{2}}{2 a^{2}}\right)\left(3 u_{1,\bf{k}}+K u_{2,\bf{k}}\right) . \\
u_{2,\bf{k}}^{\prime}=&~~ w_{2,\bf{k}}-\left(\frac{P a^{2}}{2 a^{2}}\right)\left(3 u_{2,\bf{k}}+K u_{1,\bf{k}}\right) . \\
w_{1,\bf{k}}^{\prime}=&~~-\left(m_{1}^{2}+\frac{K^{2} P_{a}^{2}}{4 a^{4}}\right) u_{1,\bf{k}}-\left(\frac{P_{a}^{2}}{2 a^{2}}\right)\left(3 w_{1,\bf{k}}+K w_{2,\bf{k}}\right) \\
&+\left(\frac{K P_{a}^{2}}{2 a^{4}}\right) u_{2,\bf{k}} .\\
w_{2,\bf{k}}^{\prime}=&~~-\left(m_{2}^{2}+\frac{K^{2} P_{a}^{2}}{4 a^{4}}\right) u_{2,\bf{k}}-\left(\frac{P_{a}^{2}}{2 a^{2}}\right)\left(3 w_{2,\bf{k}}+K w_{1,\bf{k}}\right) \\
&+\left(\frac{K P_{a}^{2}}{2 a^{4}}\right) u_{1,\bf{k}} .
\end{aligned}
\label{modediff}
\end{equation}
We can notice that if we switch off the interaction strength between the two fields by setting $K=0$, we will get two independent sectors of scalar fields which are symmetric and decoupled with each other. Now having these mode function equations are very useful because one can calculate the equation of motion for $R$, $\Theta$ and $\Phi$ for both the interacting fields. 

Before moving further we will give a summary of various calculations done till now. We have quantized the Hamiltonian of the two coupled scalar fields in de Sitter background space Eq\eqref{hamilto}, We have shown how the quantized Hamiltonian can be mapped to the two-mode coupled harmonic oscillator. We calculated the four-mode squeezed operator for this quantized Hamiltonian and using the four-mode squeezed operator and the total rotation operator we have defined the evolution operator, which have $R_{1,\bf{k}}$, $\Theta_{1,\bf{k}}$, $\Phi_{1,\bf{k}}$, $R_{2,\bf{k}}$, $\Theta_{2,\bf{k}}$ and $\Phi_{2,\bf{k}}$ as functional variables and this evolution operator governs the properties of the vacuum state for the present cosmological setup. Using the expansion of the position and momentum operators in terms of annihilation and creation operators in Heisenberg's representation we calculated the coupled differential equation for the mode functions.  From Eq\eqref{mo1} we get the time dependent position and momentum operators where, we have used the Heisenberg representation of $b_{1,\bf{k}}$ and $b_{2,\bf{k}}$.

Here, Eq\eqref{69eqt} consists of time dependent expression for the operators $\hat{v}_{1,\bf{k}}$ and $\hat{\pi}_{1,\bf{k}}$ 

\begin{widetext}
\begin{equation}
\begin{aligned}
\hat{v}_{1,\bf{k}}(\eta)=&\frac{1}{\sqrt{2 k}}\Bigg[ b_{1,\bf{k}}\left(\cosh R_{1,\bf{k}} e^{-i \Phi_{1,\bf{k}}}-\sinh R_{1,\bf{k}} e^{-i\left(\Phi_{1,\bf{k}}+2 \Theta_{1,\bf{k}}\right)}+Ke^{i \Phi_{1,\bf{k}}}\left(-\sinh R_{1,\bf{k}} e^{2 i \Theta_{1,\bf{k}}}\frac{{X_4}}{6}+\cosh R_{1,\bf{k}} \frac{{X_3}^{\dagger}}{6}\right)\right) \\
&\left.+b_{1,-\bf{k}}^{\dagger}\left(\cosh R_{1,\bf{k}} e^{i \Phi_{1,\bf{k}}}-\sinh R_{1,\bf{k}} e^{i\left(\Phi_{1,\bf{k}}+2 \Theta_{1,\bf{k}}\right)}+Ke^{i \Phi_{1,\bf{k}}}\left(-\sinh R_{1,\bf{k}} e^{2 i \Theta_{1,\bf{k}}}\frac{{{X_4}}^{\dagger}}{6}+\cosh R_{1,\bf{k}} \frac{{X_3}}{6}\right)\right)\right]\\&+\cdots\\
\hat{\pi}_{1,\vec{k}}(\eta)=&-i \sqrt{\frac{k}{2}}\Bigg[ b_{1,\bf{k}}\left(\cosh R_{1,\bf{k}} e^{-i \Phi_{1,\bf{k}}}+\sinh R_{1,\bf{k}} e^{-i\left(\Phi_{1,\bf{k}}+2 \Theta_{1,\bf{k}}\right)}-Ke^{i \Phi_{1,\bf{k}}}\left(\sinh R_{1,\bf{k}} e^{2 i \Theta_{1,\bf{k}}}\frac{{X_4}}{6}+\cosh R_{1,\bf{k}} \frac{{X_3}^{\dagger}}{6}\right)\right) \\
&\left.+b_{1,-\bf{k}}^{\dagger}\left(\cosh R_{1,\bf{k}} e^{i \Phi_{1,\bf{k}}}+\sinh R_{1,\bf{k}} e^{i\left(\Phi_{1,\bf{k}}+2 \Theta_{1,\bf{k}}\right)}+Ke^{i \Phi_{1,\bf{k}}}\left(-\sinh R_{1,\bf{k}} e^{2 i \Theta_{1,\bf{k}}}\frac{{X_4}^{\dagger}}{6}-\cosh R_{1,\bf{k}} \frac{{X_3}}{6}\right)\right)\right]\\
&+\cdots
    \end{aligned}
    \label{69eqt}
\end{equation}
\end{widetext}

Here, Eq\eqref{71eq} consists of time dependent expression for the operators $\hat{v}_{1,\bf{k}}$ and $\hat{\pi}_{1,\bf{k}}$ and the $ ``\cdots''$ in Eq(\eqref{69eqt},~\eqref{71eq}) represents the contributions from other field which we will neglect because in Eq(\eqref{mo1},\eqref{60eq}) the position and momentum operators do not contain the creation and annihilation operators of other fields.

\begin{widetext}

\begin{equation}
\begin{aligned}
\hat{v}_{2,\bf{k}}(\eta)=&\frac{1}{\sqrt{2 k}}\Bigg[ b_{2,\bf{k}}\left(\cosh R_{2,\bf{k}} e^{-i \Phi_{2,\bf{k}}}-\sinh R_{2,\bf{k}} e^{-i\left(\Phi_{2,\bf{k}}+2 \Theta_{2,\bf{k}}\right)}+Ke^{i \Phi_{2,\bf{k}}}\left(\sinh R_{2,\bf{k}} e^{2 i \Theta_{2,\bf{k}}}\frac{{X_4}}{6}-\cosh R_{2,\bf{k}} \frac{{X_3}^{\dagger}}{6}\right)\right) \\
&\left.+b_{1,-\bf{k}}^{\dagger}\left(\cosh R_{2,\bf{k}} e^{i \Phi_{2,\bf{k}}}-\sinh R_{2,\bf{k}} e^{i\left(\Phi_{2,\bf{k}}+2 \Theta_{2,\bf{k}}\right)}+Ke^{i \Phi_{2,\bf{k}}}\left(\sinh R_{2,\bf{k}} e^{2 i \Theta_{2,\bf{k}}}\frac{{X_4}^{\dagger}}{6}-\cosh R_{2,\bf{k}} \frac{X_3}{6}\right)\right)\right]
\\&+\cdots\\
\hat{\pi}_{1,\vec{k}}(\eta)=&-i \sqrt{\frac{k}{2}}\Bigg[ b_{2,\bf{k}}\left(\cosh R_{2,\bf{k}} e^{-i \Phi_{2,\bf{k}}}+\sinh R_{2,\bf{k}} e^{-i\left(\Phi_{2,\bf{k}}+2 \Theta_{2,\bf{k}}\right)}+Ke^{i \Phi_{2,\bf{k}}}\left(\sinh R_{2,\bf{k}} e^{2 i \Theta_{2,\bf{k}}}\frac{{X_4}}{6}+\cosh R_{2,\bf{k}} \frac{{X_3}^{\dagger}}{6}\right)\right) \\
&\left.+b_{1,-\bf{k}}^{\dagger}\left(\cosh R_{2,\bf{k}} e^{i \Phi_{2,\bf{k}}}+\sinh R_{2,\bf{k}} e^{i\left(\Phi_{2,\bf{k}}+2 \Theta_{2,\bf{k}}\right)}+Ke^{i \Phi_{2,\bf{k}}}\left(\sinh R_{2,\bf{k}} e^{2 i \Theta_{2,\bf{k}}}\frac{{X_4}^{\dagger}}{6}+\cosh R_{2,\bf{k}} \frac{X_3}{6}\right)\right)\right]
\\&+\cdots
    \end{aligned}
    \label{71eq}
\end{equation}
\end{widetext}

On comparing the time dependent form of operators for position and momentum for both the fields given in Eq\eqref{69eqt},\eqref{71eq} with \eqref{mo1} we can identify the mode functions to be 

\onecolumngrid
\begin{equation}
\begin{aligned}
&u_{1,\bf{k}}(\eta)=\frac{1}{\sqrt{2 k}}\left[\cosh R_{1,\bf{k}} e^{i \Phi_{1,\bf{k}}}-\sinh R_{1,\bf{k}} e^{i (\Phi_{1,\bf{k}}+2\Theta_{1,\bf{k}})}+Ke^{i \Phi_{1,\bf{k}}}\left(-\sinh R_{1,\bf{k}} e^{2 i \Theta_{1,\bf{k}}}\frac{{X_4}^{\dagger}}{6}+\cosh R_{1,\bf{k}} \frac{X_3}{6}\right)\right] \\
&w_{1,\bf{k}}(\eta)=i\sqrt{\frac{k}{2}}\left[\cosh R_{1,\bf{k}} e^{i \Phi_{1,\bf{k}}}+\sinh R_{1,\bf{k}} e^{i (\Phi_{1,\bf{k}}+2\Theta_{1,\bf{k}})}+Ke^{i \Phi_{1,\bf{k}}}\left(-\sinh R_{1,\bf{k}} e^{2 i \Theta_{1,\bf{k}}}\frac{{X_4}^{\dagger}}{6}-\cosh R_{1,\bf{k}} \frac{X_3}{6}\right)\right] \\
&u_{2,\bf{k}}(\eta)=\frac{1}{\sqrt{2 k}}\left[\cosh R_{2,\bf{k}} e^{i \Phi_{2,\bf{k}}}-\sinh R_{2,\bf{k}} e^{i (\Phi_{2,\bf{k}}+2\Theta_{2,\bf{k}})}+Ke^{i \Phi_{2,\bf{k}}}\left(\sinh R_{2,\bf{k}} e^{2 i \Theta_{2,\bf{k}}}\frac{{X_4}^{\dagger}}{6}-\cosh R_{2,\bf{k}} \frac{X_3}{6}\right)\right] \\
&w_{2,\bf{k}}(\eta)=i\sqrt{\frac{k}{2}}\left[\cosh R_{2,\bf{k}} e^{i \Phi_{2,\bf{k}}}+\sinh R_{2,\bf{k}} e^{i (\Phi_{2,\bf{k}}+2\Theta_{2,\bf{k}})}+Ke^{i \Phi_{2,\bf{k}}}\left(\sinh R_{2,\bf{k}} e^{2 i \Theta_{2,\bf{k}}}\frac{{X_4}^{\dagger}}{6}+\cosh R_{2,\bf{k}} \frac{X_3}{6}\right)\right]
\end{aligned}
\end{equation}

and these above equations define the transformation between the variables which are in the Schr\"{o}dinger representation with the mode functions in the Heisenberg representation. It is now a matter of algebra to show that Hamilton’s equations for the mode functions
\eqref{modediff} give the equations of motion for

\begin{equation}\label{76eqt}
\begin{aligned}
R_{1,\bf{k}}^{\prime} &=\lambda_{1,\bf{k}} \cos 2\left(\varphi_{1,\bf{k}}-\Theta_{1,\bf{k}}\right) + K{Y_1}\\ \Theta_{1,\bf{k}}^{\prime} &=-\Omega_{1,\bf{k}}+\frac{\lambda_{1,\bf{k}}}{2}\left(\tanh R_{1,\bf{k}}+\operatorname{coth} R_{1,\bf{k}}\right) \sin 2\left(\varphi_{1,\bf{k}}-\Theta_{1,\bf{k}}\right) \\ &+ K{Y_2} \\ \Phi_{1,\bf{k}}^{\prime} &=\Omega_{1,\bf{k}}-\lambda_{1,\bf{k}} \tanh R_{1,\bf{k}} \sin 2\left(\varphi_{1,\bf{k}}-\Theta_{1,\bf{k}}\right) + K{Y_3}\\
R_{2,\bf{k}}^{\prime} &=\lambda_{2,\bf{k}} \cos 2\left(\varphi_{2,\bf{k}}-\Theta_{2,\bf{k}}\right) + K{Y_4} \\ \Theta_{2,\bf{k}}^{\prime} &=-\Omega_{2,\bf{k}}+\frac{\lambda_{2,\bf{k}}}{2}\left(\tanh R_{2,\bf{k}}+\operatorname{coth} R_{2,\bf{k}}\right) \sin 2\left(\varphi_{2,\bf{k}}-\Theta_{2,\bf{k}}\right) \\ &+ K{Y_5} \\ \Phi_{2,\bf{k}}^{\prime} &=\Omega_{2,\bf{k}}-\lambda_{2,\bf{k}} \tanh R_{2,\bf{k}} \sin 2\left(\varphi_{2,\bf{k}}-\Theta_{2,\bf{k}}\right) + K{Y_6}
\end{aligned}
\end{equation}

\twocolumngrid
Where $Y_i$'s are the contributions coming from the interaction between the two  scalar fields. Here we have considered the effects of perturbation terms only upto $\mathcal{O}(R^2$). $Y_i$'s are given in the appendix. \ref{sec:appendixA}. 

and

\begin{align}
&\Omega_{i,\bf{k}}=\frac{k}{2}\left(1+c_{si}^{2}\right) \\
&\lambda_{i,\bf{k}}=\left[\left(\frac{k}{2}\left(1-c_{si}^{2}\right)\right)^{2}+\left(\frac{{3P_a}^{2}}{2a^2}\right)^{2}\right]^{\frac{1}{2}} \\
&\varphi_{i,\bf{k}}=-\frac{\pi}{2}+\frac{1}{2} \arctan \left(\frac{k a^2}{3 {P_a}^{2}}\left(1-c_{si}^{2}\right)\right) .
\end{align}

The index $i$ runs from 1 and 2. Here the $c_{si}$ is the effective sound speed for two fields and is given by

\begin{equation}
\begin{aligned}
k^2~c_{si}^2 =\left[\left({m_i}\right)^{2}+\left(\frac{{{K^2}P_a}^{2}}{4a^4}\right)\right]
\end{aligned}
\end{equation}
Note: We will not consider term with $K^2$ because we are considering terms which are first order in  $K$.

The differential equations given in Eq[\eqref{76eqt}] are very useful because these equations govern the dynamics of the variables: $R_{1,\bf{k}}$, $\Theta_{1,\bf{k}}$, $\Phi_{1,\bf{k}}$, $R_{2,\bf{k}}$, $\Theta_{2,\bf{k}}$ and $\Phi_{2,\bf{k}}$ which effects the evolution operator for the four mode squeezed states in de Sitter background space. Hence, these equations can be used for numerical studies of the two coupled scalar fields in de Sitter space.


\section{\textcolor{Sepia}{\textbf{ \Large Numerical Analysis}}}\label{Numeri}
In this Section we present the numerical analysis for two coupled scalar fields which are weakly interacting with coupling strength $K\ll 1$ and we will do the numerical analysis in the limit where the interaction terms are highly linear in nature. We will study the behaviour of squeezing parameter with respect to the conformal time for this particular setup.
We give the differential equation for squeezed parameters of the two coupled scalar fields in de Sitter background space in eq \eqref{linear_eqt}. The interacting parts in these set of six equations contain only terms which are linear in nature. We have chosen this limit in order to understand the numerical behaviour of squeezing parameters in a compact manner. For numerical analysis we set the momentum $p_{a}=0.1$, The Hubble parameter is set to 0.1 in natural units, we take $k=0.0001 {\rm Mpc}^{-1}$ and $a=-\frac{1}{H\eta}$.
 
\begin{widetext}

\begin{equation}\label{linear_eqt}
\begin{aligned}
R_{1,\bf{k}}^{\prime}(\eta)&=\lambda_{1,\bf{k}}(\eta) \operatorname{cos}[2(\varphi_{1,\bf{k}}(\eta)+\Theta_{1,\bf{k}}(\eta))]+\frac{K}{{6 k}}\bigg(\mathrm{i}(2 k P(\eta)  R_{1,\bf{k}}(\eta)+(-2 k P(\eta)+3 \mathrm{i} S(\eta)) R_{2,\bf{k}}(\eta)-3 S(\eta)(\mathrm{i}+\Phi_{1,\bf{k}}(\eta)\\
&\left.-\Phi_{2,\bf{k}}(\eta))\bigg)\right)\\
R_{2,\bf{k}}^{\prime}(\eta)&=\lambda_{2,\bf{k}}(\eta) \operatorname{cos}[2(\varphi_{2,\bf{k}}(\eta)+\Theta_{1,\bf{k}}(\eta))]+K\left(\frac{1}{3}\left(P(\eta)-P(\eta)  R_{1,\bf{k}}(\eta)+P(\eta)  R_{2,\bf{k}}(\eta)+\frac{3 S(\eta)  \Theta_{2,\bf{k}}(\eta)}{k}\right)\right)\\
\Theta_{1,\bf{k}}^{\prime}(\eta)&=-\Omega_{1,\bf{k}}-(\lambda_{1,\bf{k}} \operatorname{sin}[2(\varphi_{1,\bf{k}}+\Theta_{1,\bf{k}}(\eta))]) \operatorname{coth}[2 R_{1,\bf{k}}(\eta)]+
K\left(\frac{1}{12 kR_{1,\bf{k}}(\eta)}\left(3 S(\eta)+(-2 i k P(\eta)-3 S(\eta)) R_{2,\bf{k}}(\eta)\right.\right.\\
&~~~\left.\left.-4 k P(\eta)  \Theta_{1,\bf{k}}(\eta)-2 k P(\eta)  \Phi_{1,\bf{k}}(\eta)-3 \mathrm{i} S(\eta)  \Phi_{1,\bf{k}}(\eta)+2 k P(\eta)  \Phi_{2,\bf{k}}(\eta)+3 \mathrm{i} S(\eta)  \Phi_{2,\bf{k}}(\eta)+\right.\right.
R_{1,\bf{k}}(\eta)(2 \mathrm{i} kP(\eta)-3 S(\eta)\\
&-4 kP(\eta) \Theta_{1,\bf{k}} (\eta)+6 i S(\eta) \Theta_{1,\bf{k}} (\eta)+R_{2,\bf{k}}(\eta)(3 S(\eta)+8 kP(\eta) \Theta_{1,\bf{k}}(\eta))+3 i S(\eta) \Phi_{1,\bf{k}}(\eta)-3 i S(\eta) \Phi_{2,\bf{k}}(\eta)))\bigg)\\
\Theta_{2,\bf{k}}^{\prime}(\eta)&=-\Omega_{2,\bf{k}}-(\lambda_{2,\bf{k}} \operatorname{sin}[2(\varphi_{2,\bf{k}}+\Theta_{2,\bf{k}}(\eta))]) \operatorname{coth}[2 R_{2,\bf{k}}(\eta)]+
K\bigg(\frac{1}{12 k R_{2,\bf{k}}(\eta)}(3 S(\eta)+R_{1,\bf{k}}(\eta)(-2 \mathrm{i} k P(\eta)-3 S(\eta)+\\
&3 S(\eta) R_{2,\bf{k}}(\eta))-4 kP(\eta)\Theta_{2,\bf{k}}(\eta)+2 kP(\eta) \Phi_{1,\bf{k}}(\eta)+3 \mathrm{i} \mathrm{S}(\eta) \Phi_{1,\bf{k}}(\eta)-2 kP(\eta) \Phi_{2,\bf{k}}(\eta)-
3 i S(\eta) \Phi_{2,\bf{k}}(\eta)\\
&+R_{2,\bf{k}}(\eta)(2 i k P(\eta)-3 S(\eta)-4 k P(\eta) \Theta_{2,\bf{k}}(\eta)+6 i S(\eta) \Theta_{2,\bf{k}}(\eta)-3 i S(\eta) \Phi_{1,\bf{k}}(\eta)+3 i S(\eta) \Phi_{2,\bf{k}}(\eta)))\bigg)\\
\Phi_{1,\bf{k}}^{\prime}(\eta)&=\Omega_{1,\bf{k}}-\lambda_{1,\bf{k}}\operatorname{tanh}[R_{1,\bf{k}}(\eta)] \operatorname{sin} [2(\varphi_{1,\bf{k}}+\Theta_{1,\bf{k}}(\eta))]\\
&+K\left(\frac{i(2 k P(\eta)  R_{1,\bf{k}}(\eta)+(-2 k P(\eta)+3 iS(\eta)) R_{2,\bf{k}}(\eta)-3 S(\eta)(i+\Phi_{1,\bf{k}}(\eta)-\Phi_{2,\bf{k}}(\eta)))}{6 k}\right)\\
\Phi_{2,\bf{k}}^{\prime}(\eta)&=\Omega_{2,\bf{k}}-\lambda_{2,\bf{k}}\operatorname{tanh}[R_{2,\bf{k}}(\eta)] \operatorname{sin} [2(\varphi_{2,\bf{k}}+\Theta_{2,\bf{k}}(\eta))]\\
&-K\left(\frac{{i}((2 k P(\eta)-3 i S(\eta)) R_{1,\bf{k}}(\eta)-2 k P(\eta)  R_{2,\bf{k}}(\eta)+3 S(\eta)(i-\Phi_{1,\bf{k}}(\eta)+\Phi_{2,\bf{k}}(\eta)))}{6 k}\right)\\
\end{aligned}
\end{equation}

\end{widetext}

\begin{figure*}[htb!]
	\centering
	\subfigure{
		\includegraphics[width=8cm] {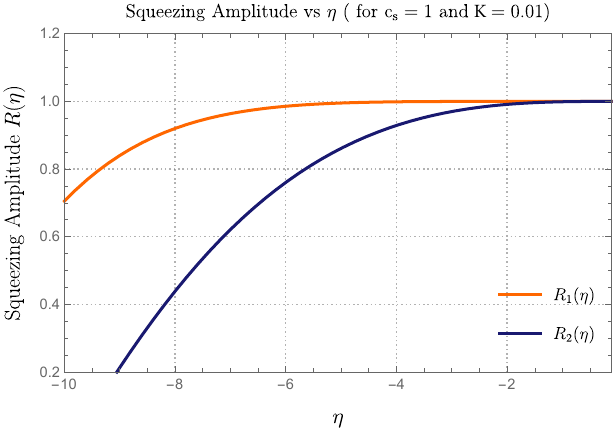}\label{entvsacs1}
	}
	\subfigure{
		\includegraphics[width=8.2cm] {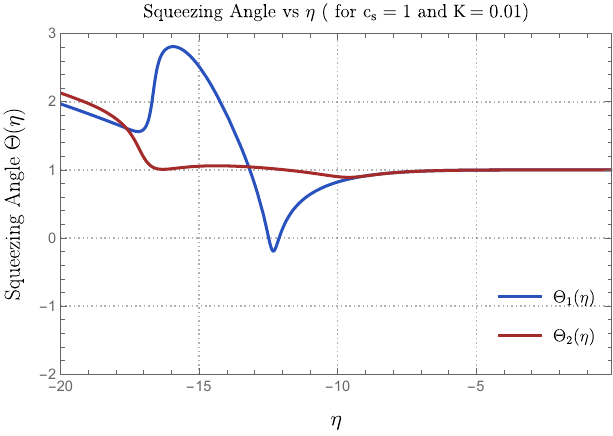}\label{entvsacs2}
	}
		\subfigure{
		\includegraphics[width=8.2cm] {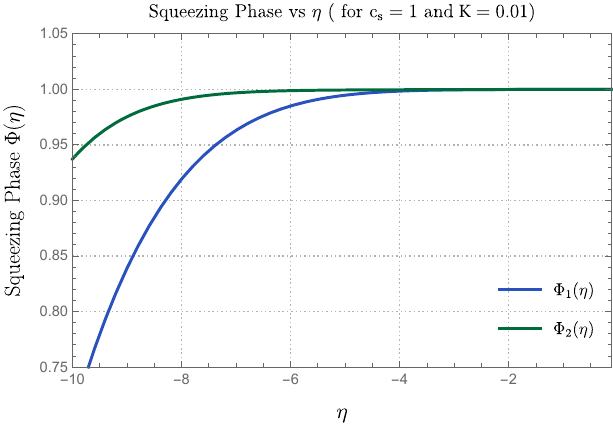}\label{entvsacs2}
	}
	\caption{Behaviour of the squeezing parameters, such as squeezing amplitude in fig (2(a)), squeezing angle in fig (2(b)), squeezing phase in fig (2(c)) with respect to the conformal time $\eta$ for two coupled scalar fields with coupling strength $K=0.01$ and the effective sound speed parameter $c_{s1}$ and $c_{s2}$ for both the fields is equal to 1.}
	\label{entvsacs}
\end{figure*}

In fig (1), we have plotted three graphs which shows the behaviour of squeezing parameters with respect to the $\eta$ for the two coupled scalar fields Fig (1(a)) represents behaviour of the squeezing amplitude $R_{1,\bf{k}}$ for first field and $R_{2,\bf{k}}$ for second field with respect to the conformal time $\eta$. We observe that the squeezing amplitude at early times for the two fields are different which is due to the reason that the contribution from the interacting terms in both the field is different. We see that the $R_{1,\bf{k}}$ dominates over $R_{2,\bf{k}}$ at early times and behaviour for both the squeezing amplitude is similar i.e, both the squeezing amplitudes decrease as we move towards the more negative value of the conformal time $\eta$ and at $\eta \approx 0$ which is the present time of the universe the distinguishability between the squeezing amplitude for the two fields vanishes.  

In fig (1(b)), we have plotted the squeezing angle or rotation angle, $\Theta_{1,\bf{k}}$ for first field and $\Theta_{2,\bf{k}}$ for second field. The squeezing angle for the two fields at early time or more negative value of conformal time shows that squeezing angle of second field is dominating over the first field but as we move towards the right of the graph the field one dominates and decreases and then the second field dominates. But when we move toward the present time $\eta=0$ we see that the these interchanging dominance effects vanishes and squeezing angles for both fields saturates to a particular constant value.

The behaviour for the squeezing phase parameters with respect to the conformal time $\eta$ has been shown in fig (1(c)). Here squeezing phase $\Phi_{1,\bf{k}}$ is for first field and $\Phi_{2,\bf{k}}$ is for second field. We observe that its nature is similar to that of squeezing amplitude parameter shown in fig(1(a)), but there is one difference which is the dominance of squeezing phase of second field over the first field, which was another way around in case of squeezing amplitude parameter. We see that in case of squeezing phase the difference in the two fields vanishes and attains a constant value.


\begin{figure*}[htb!]
	\centering
	\subfigure{
		\includegraphics[width=8cm] {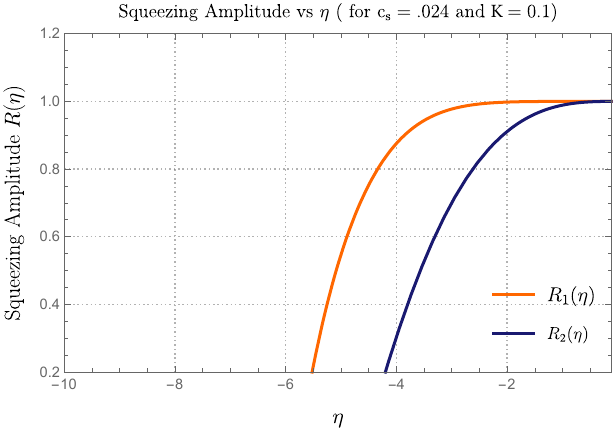}\label{entvsacs1}
	}
	\subfigure{
		\includegraphics[width=8.2cm] {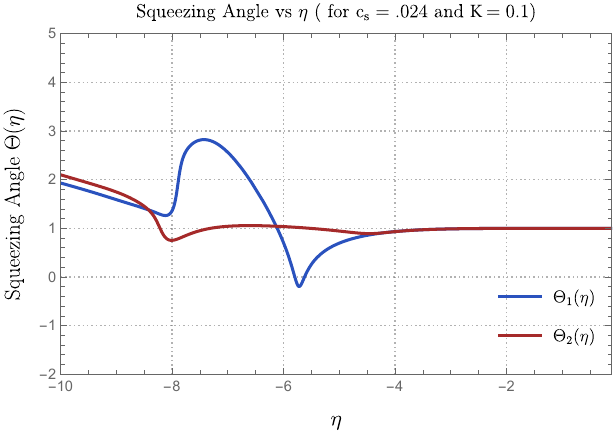}\label{entvsacs2}
	}
		\subfigure{
		\includegraphics[width=8.2cm] {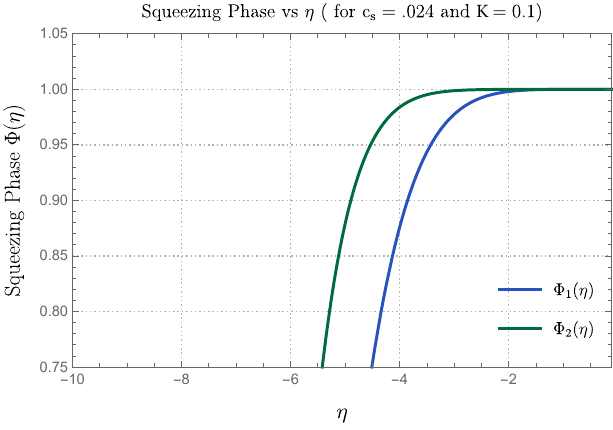}\label{entvsacs2}
	}
	\caption{Behaviour of the squeezing parameters, such as squeezing amplitude in fig (1(a)), squeezing angle in fig (1(b)), squeezing phase in fig (1(c)) with respect to the conformal time $\eta$ for two coupled scalar fields with coupling strength $K=0.1$ and the effective sound speed parameter $c_{s1}$ and $c_{s2}$ for both the fields is equal to 0.024.}
	\label{entvsacs}
\end{figure*}
In fig (2) we again plot the squeezing parameters but with a different value of effective sound speed parameter $c_{s1}$ and $c_{s2}$. We set the effective sound speed parameter for both fields to its lowest bound which is 0.024. Here the coupling strength $K=0.1$.
We observe a similar behaviour for all the three squeezing parameters but one new observation which we can see in fig (2(a)) and fig(2(c)) is that the squeezing amplitude and squeezing phase tends to zero as we move towards more negative value of conformal time. All the three graphs in fig (2) gets shifted towards right or near the origin as compared to the graphs in fig (1).

\section{\textcolor{Sepia}{\textbf{ \Large Conclusion}}}\label{CC}

This work dealt with the mathematical formalism of four mode squeezed state in cosmology. From the cosmological perspective we have encountered two scalar fields, where the metric of the background space was set to de Sitter. The motivation for choosing this particular metric was the following setup of the action which we have chosen can be helpful to understand the dynamics of the two  scalar fields in FRW cosmological universe.
The analysis of our present work can be helpful in the limit where the two scalar fields are weakly coupled $K<<1$ and the perturbation effects was taken only upto linear order $\mathcal{O}(R$). 
\\
This work could be summarised in following points:
\begin{itemize}
    \item We have quantized the modes of the two coupled scalar fields $\mu_1$ and $\mu_2$. We have also calculated the position and momentum variables for the same in de Sitter background space and obtained the quantized Hamiltonian $H$. 
    We made connection between the two coupled inverted quantum harmonic oscillator system and four mode squeezed state formalism. 
    \item We have given a detailed calculation for constructing the four mode squeezed state operator which is useful for understanding the cosmological action for the two interacting  scalar fields and also for other systems which can be explained in terms of two coupled inverted quantum harmonic oscillators.
    \item The time evolution operator for the four mode squeezed states is also given, which we have used to calculate the time dependent (Heisenberg picture) annihilation and creation operators for two coupled scalar fields in de Sitter background space. Using the Heisenberg equation of motion, we have calculated the coupled differential equation for the mode functions of the two coupled scalar fields in de Sitter space.
    \item We presented the expression for $R_{1,\bf{k}}$, $\Theta_{1,\bf{k}}$, $\Phi_{1,\bf{k}}$, $R_{2,\bf{k}}$, $\Theta_{2,\bf{k}}$ and $\Phi_{2,\bf{k}}$ which are the parameters of the evolution operator for four mode squeezed state and which governs the evolution of the state for two coupled scalar fields in de Sitter metric.
    \item We conclude the work by analysing the behaviour of the squeezing parameters for the two coupled scalar fields. We found during early conformal time, the distinction between the squeezing parameters were noticeable. As we move to current time, $\eta\approx0$, we find that the parameters converge and become indistinguishable. 
\end{itemize}

  With this tools in hand it would be interesting to compute quantum information quantities such as entanglement entropy, quantum discord, circuit complexity and many more quantum information theoretic measures for two coupled scalar fields in de Sitter space.  This will going to help us to know about the feature and the behaviour of the long range quantum correlations for the system under consideration.  Earlier the formalism was not developed for the four-mode squeezed states for which these crucial aspects was not studied in the previous works.  Now since the formalism is developed and we know how to handle the system numerically,  it would be really good to study the mentioned aspects in near future.

\textbf{Acknowledgement:}
~~~The Visiting Post Doctoral research fellowship of SC is supported by the J.  C.  Bose National Fellowship of Director, Professor Rajesh Gopakumar,  ICTS, TIFR, Bengaluru.  The research of SP is supported by the J.  C.  Bose National Fellowship.  SC also would like to thank ICTS, TIFR, Bengaluru for providing the work friendly environment. SC also would like to thank all the members of our newly formed virtual international non-profit consortium Quantum Structures of the Space-Time \& Matter (QASTM) for for elaborative discussions.  AR and NP would like to thank the members of the QASTM Forum for useful discussions.  Last but not least, we would like to acknowledge our debt to the people belonging to the various part of the
world for their generous and steady support for research
in natural sciences.

\widetext
\section{Appendix}
	\textcolor{Sepia}{\subsection{\sffamily Interacting part of differential equations}\label{sec:appendixA}}

$Y_{1}=$

$\left(-\left(A_{3} B_{2} E_{1}-A_{2} B_{3} E_{1}-A_{3} B_{1} E_{2}+A_{1} B_{3} E_{2}+A_{2} B_{1} E_{3}-A_{1} B_{2} E_{3}\right)\right. (C_{6} D_{5} F_{4}-C_{5} D_{6} F_{4}-C_{6} D_{4} F_{5}+C_{4} D_{6} F_{5}$

$+C_{5} D_{4} F_{6}-C_{4} D_{5} F_{6}) (B_{2} E_{1} a_{0}-B_{1} E_{2} a_{0}-A_{2} E_{1} b_{0}+A_{1} E_{2} b_{0}+A_{2} B_{1} e_{0}-A_{1} B_{2} e_{0})+ (x_{21} A_{2} B_{1}-x_{21} A_{1} B_{2}-x_{0} A_{2} E_{1}$

$+x_{3} B_{2} E_{1}+x_{0} A_{1} E_{2}-x_{3} B_{1} E_{2}) (A_{3} B_{2} E_{1}-A_{2} B_{3} E_{1}-A_{3} B_{1} E_{2}+A_{1} B_{3} E_{2}+A_{2} B_{1} E_{3}-A_{1} B_{2} E_{3}) (D_{6} F_{5} C_{0}-D_{5} F_{6} C_{0}$

$-C_{6} F_{5} D_{0}+C_{5} F_{6} D_{0}+C_{6} D_{5} F_{0}-C_{5} D_{6} F_{0}) R_{1}{ }^{2}+ (C_{6} D_{5} F_{4}-C_{5} D_{6} F_{4}-C_{6} D_{4} F_{5}+C_{4} D_{6} F_{5}+C_{5} D_{4} F_{6}-C_{4} D_{5} F_{6})$

$\left(B_{3} E_{2} A_{0}-B_{2} E_{3} A_{0}-A_{3} E_{2} B_{0}+A_{2} E_{3} B_{0}+A_{3} B_{2} E_{0}-A_{2} B_{3} E_{0}\right) R_{2} (-x_{7} A_{2} E_{1}+x_{7} A_{1} E_{2}+ (-x_{18} A_{2} B_{1}+x_{18} A_{1} B_{2}$

$-x_{6} A_{2} E_{1}+x_{2} B_{2} E_{1}+x_{6} A_{1} E_{2}-x_{2} B_{1} E_{2}) R_{2}) R_{1} (x_{5}(A_{2} E_{1}-A_{1} E_{2}) (C_{6} D_{5} F_{4}-C_{5} D_{6} F_{4}-C_{6} D_{4} F_{5}+C_{4} D_{6} F_{5}+C_{5} D_{4} F_{6}$

$-C_{4} D_{5} F_{6}) (-B_{3} E_{2} A_{0}+B_{2} E_{3} A_{0}+A_{3} E_{2} B_{0}-A_{2} E_{3} B_{0}-A_{3} B_{2} E_{0}+A_{2} B_{3} E_{0})+ (-x_{19}\left(A_{2} B_{1}-A_{1} B_{2}\right)(C_{6} D_{5} F_{4}$

$-C_{5} D_{6} F_{4}-C_{6} D_{4} F_{5}+C_{4} D_{6} F_{5}+C_{5} D_{4} F_{6}-C_{4} D_{5} F_{6}) (-B_{3} E_{2} A_{0}+B_{2} E_{3} A_{0}+A_{3} E_{2} B_{0}-A_{2} E_{3} B_{0}-A_{3} B_{2} E_{0}$

$+A_{2} B_{3} E_{0})- x_{1} (B_{2} E_{1}-B_{1} E_{2}) (C_{6} D_{5} F_{4}-C_{5} D_{6} F_{4}-C_{6} D_{4} F_{5}+C_{4} D_{6} F_{5}+C_{5} D_{4} F_{6}-C_{4} D_{5} F_{6}) (-B_{3} E_{2} A_{0} +B_{2} E_{3} A_{0}$

$+A_{3} E_{2} B_{0}-A_{2} E_{3} B_{0}-A_{3} B_{2} E_{0} +A_{2} B_{3} E_{0})- (x_{20} A_{2} B_{1}-X_{20} A_{1} B_{2}+ x_{8} A_{2} E_{1}-x_{4} B_{2} E_{1}-x_{8} A_{1} E_{2} +x_{4} B_{1} E_{2})$

$~(A_{3} B_{2} E_{1}-A_{2} B_{3} E_{1}-A_{3} B_{1} E_{2}+A_{1} B_{3} E_{2} + A_{2} B_{1} E_{3}-A_{1} B_{2} E_{3}) (D_{6} F_{5} C_{0}-D_{5} F_{6} C_{0}-C_{6} F_{5} D_{0} +C_{5} F_{6} D_{0}+C_{6} D_{5} F_{0}$

$\left.-C_{5} D_{6} F_{0})) R_{2})\right) / ((A_{3} B_{2} E_{1}-A_{2} B_{3} E_{1}-A_{3} B_{1} E_{2}+A_{1} B_{3} E_{2}+A_{2} B_{1} E_{3}-A_{1} B_{2} E_{3})^{2} (C_{6} D_{5} F_{4}-C_{5} D_{6} F_{4}-C_{6} D_{4} F_{5}$

$+C_{4} D_{6} F_{5}+C_{5} D_{4} F_{6}-C_{4} D_{5} F_{6}))$

$Y_{4}=$

$((-C_{6} D_{5} F_{4}+C_{5} D_{6} F_{4}+C_{6} D_{4} F_{5}-C_{4} D_{6} F_{5}-C_{5} D_{4} F_{6}+C_{4} D_{5} F_{6}) (B_{3}((-A_{2} E_{1}+A_{1} E_{2})(D_{6} F_{5}-D_{5} F_{6}) c_{0}+$

$C_{6} d_{1} F_{5}(E_{2} A_{0}-A_{2} E_{0})+C_{5} d_{1} F_{6}(-E_{2} A_{0}+A_{2} E_{0})+ C_{6} (A_{2} E_{1}-A_{1} E_{2}) (F_{5} d_{0}-D_{5} f_{0})-C_{5}(A_{2} E_{1}-A_{1} E_{2})$

$(F_{6} d_{0}-D_{6} f_{0}))+ B_{2}((A_{3} E_{1}-A_{1} E_{3})(D_{6} F_{5}-D_{5} F_{6}) c_{0}+C_{5} d_{1} F_{6}(E_{3} A_{0}-A_{3} E_{0})+C_{6} d_{1} F_{5}(-E_{3} A_{0}+A_{3} E_{0})-$

$ C_{6}(A_{3} E_{1}-A_{1} E_{3})(F_{5} d_{0}-D_{5} f_{0})+C_{5}(A_{3} E_{1}-A_{1} E_{3})(F_{6} d_{0}-D_{6} f_{0}))+ (A_{3} E_{2}-A_{2} E_{3})(B_{1}(-D_{6} F_{5}+D_{5} F_{6}) c_{0}$

$-C_{6}(d_{1} F_{5} B_{0}-B_{1} F_{5} d_{0}+B_{1} D_{5} f_{0})+ C_{5}(d_{1} F_{6} B_{0}-B_{1} F_{6} d_{0}+B_{1} D_{6} f_{0})))+ (A_{3} B_{2} E_{1}-A_{2} B_{3} E_{1}-A_{3} B_{1} E_{2}$

$+A_{1} B_{3} E_{2}+A_{2} B_{1} E_{3}-A_{1} B_{2} E_{3}) (x_{25} C_{6} D_{5}-x_{25} C_{5} D_{6}-x_{17} C_{6} F_{5}+x_{12} D_{6} F_{5}+x_{17} C_{5} F_{6}-x_{12} D_{5} F_{6}) (D_{6} F_{5} C_{0}$

$-D_{5} F_{6} C_{0}-C_{6} F_{5} D_{0}+C_{5} F_{6} D_{0}+C_{6} D_{5} F_{0}-C_{5} D_{6} F_{0}) R_{1}^{2}+ (x_{23}(C_{6} D_{5}-C_{5} D_{6})(C_{6} D_{5} F_{4}-C_{5} D_{6} F_{4}-C_{6} D_{4} F_{5}$

$+C_{4} D_{6} F_{5}+C_{5} D_{4} F_{6}-C_{4} D_{5} F_{6}) (B_{3} E_{2} A_{0}-B_{2} E_{3} A_{0}-A_{3} E_{2} B_{0}+A_{2} E_{3} B_{0}+A_{3} B_{2} E_{0}-A_{2} B_{3} E_{0})+ x_{11}(D_{6} F_{5}-D_{5} F_{6})$

$(C_{6} D_{5} F_{4}-C_{5} D_{6} F_{4}-C_{6} D_{4} F_{5}+C_{4} D_{6} F_{5}+C_{5} D_{4} F_{6}-C_{4} D_{5} F_{6}) (B_{3} E_{2} A_{0}-B_{2} E_{3} A_{0}-A_{3} E_{2} B_{0}+A_{2} E_{3} B_{0}+A_{3} B_{2} E_{0}$

$-A_{2} B_{3} E_{0})- (A_{3} B_{2} E_{1}-A_{2} B_{3} E_{1}-A_{3} B_{1} E_{2}+A_{1} B_{3} E_{2}+A_{2} B_{1} E_{3}-A_{1} B_{2} E_{3}) (x_{26} C_{6} D_{5}-x_{26} C_{5} D_{6} +x_{16} C_{6} F_{5}$

$-x_{13} D_{6} F_{5}-x_{16} C_{5} F_{6}+x_{13} D_{5} F_{6}) (D_{6} F_{5} C_{0}-D_{5} F_{6} C_{0}-C_{6} F_{5} D_{0}+C_{5} F_{6} D_{0}+C_{6} D_{5} F_{0}-C_{5} D_{6} F_{0}))R_1 R_2$

$+(x_{24} C_{6} D_{5}-x_{24} C_{5} D_{6}+x_{10} D_{6} F_{5}-x_{10} D_{5} F_{6}) (C_{6} D_{5} F_{4}-C_{5} D_{6} F_{4}-C_{6} D_{4} F_{5}+C_{4} D_{6} F_{5}+C_{5} D_{4} F_{6}-C_{4} D_{5} F_{6})$

$~(B_{3} E_{2} A_{0}-B_{2} E_{3} A_{0}-A_{3} E_{2} B_{0}+A_{2} E_{3} B_{0}+A_{3} B_{2} E_{0}-A_{2} B_{3} E_{0}) R_2^{2}) / ((A_{3} B_{2} E_{1}-A_{2} B_{3} E_{1}-A_{3} B_{1} E_{2}$

$+A_{1} B_{3} E_{2}+A_{2} B_{1} E_{3}-A_{1} B_{2} E_{3}) (C_{6} D_{5} F_{4}-C_{5} D_{6} F_{4}-C_{6} D_{4} F_{5}+C_{4} D_{6} F_{5}+C_{5} D_{4} F_{6}-C_{4} D_{5} F_{6})^{2})$

$Y_{2}=$

$((A_{3} B_{2} E_{1}-A_{2} B_{3} E_{1}-A_{3} B_{1} E_{2}+A_{1} B_{3} E_{2}+A_{2} B_{1} E_{3}-A_{1} B_{2} E_{3})(B_{3} E_{1} a_{0}-B_{1} E_{3} a_{0}-A_{3} E_{1} b_{0}+A_{1} E_{3} b_{0}$

$+A_{3} B_{1} e_{0}-A_{1} B_{3} e_{0})+ ((x_{21} A_{3} B_{1}-x_{21} A_{1} B_{3}-x_{0} A_{3} E_{1}+x_{3} B_{3} E_{1}+x_{0} A_{1} E_{3}-x_{3} B_{1} E_{3})(-A_{3} B_{2} E_{1}+A_{2} B_{3} E_{1}$

$+A_{3} B_{1} E_{2}-A_{1} B_{3} E_{2}-A_{2} B_{1} E_{3}+A_{1} B_{2} E_{3})(D_{6} F_{5} C_{0}-D_{5} F_{6} C_{0}-C_{6} F_{5} D_{0}+C_{5} F_{6} D_{0}+C_{6} D_{5} F_{0}-C_{5} D_{6} F_{0}) R_1^{2}) /$

$(C_{6} D_{5} F_{4}-C_{5} D_{6} F_{4}-C_{6} D_{4} F_{5}+C_{4} D_{6} F_{5}+C_{5} D_{4} F_{6}-C_{4} D_{5} F_{6})+x_{7}(A_{3} E_{1}-A_{1} E_{3})(B_{3} E_{2} A_{0}-B_{2} E_{3} A_{0}-A_{3} E_{2} B_{0}$

$+A_{2} E_{3} B_{0}+A_{3} B_{2} E_{0}-A_{2} B_{3} E_{0}) R_2+ (x_{18} A_{3} B_{1}-x_{18} A_{1} B_{3}+x_{6} A_{3} E_{1}-x_{2} B_{3} E_{1}-x_{6} A_{1} E_{3}+x_{2} B_{1} E_{3})(B_{3} E_{2} A_{0}-B_{2} E_{3} A_{0}$

$-A_{3} E_{2} B_{0}+A_{2} E_{3} B_{0}+A_{3} B_{2} E_{0}-A_{2} B_{3} E_{0}) R_2^{2}+ R_1(x_{5}(A_{3} E_{1}-A_{1} E_{3})(B_{3} E_{2} A_{0}-B_{2} E_{3} A_{0}-A_{3} E_{2} B_{0}+A_{2} E_{3} B_{0}$

$+A_{3} B_{2} E_{0} -A_{2} B_{3} E_{0})+ (-x_{1}(B_{3} E_{1}-B_{1} E_{3})(B_{3} E_{2} A_{0}-B_{2} E_{3} A_{0}-A_{3} E_{2} B_{0}+A_{2} E_{3} B_{0}+A_{3} B_{2} E_{0}-A_{2} B_{3} E_{0})$

$+x_{19}(A_{3} B_{1}-A_{1} B_{3}) (-B_{3} E_{2} A_{0}+B_{2} E_{3} A_{0}+A_{3} E_{2} B_{0}-A_{2} E_{3} B_{0}-A_{3} B_{2} E_{0}+A_{2} B_{3} E_{0})- ((x_{20} A_{3} B_{1}-x_{20} A_{1} B_{3}$

$+x_{8} A_{3} E_{1}-x_{4} B_{3} E_{1} -x_{8} A_{1} E_{3}+x_{4} B_{1} E_{3})(-A_{3} B_{2} E_{1}+A_{2} B_{3} E_{1}+A_{3} B_{1} E_{2}-A_{1} B_{3} E_{2}-A_{2} B_{1} E_{3}+A_{1} B_{2} E_{3})$

$(D_{6} F_{5} C_{0}-D_{5} F_{6} C_{0}-C_{6} F_{5} D_{0}+C_{5} F_{6} D_{0}+C_{6} D_{5} F_{0}-C_{5} D_{6} F_{0})) / (C_{6} D_{5} F_{4}-C_{5} D_{6} F_{4}-C_{6} D_{4} F_{5}+C_{4} D_{6} F_{5}$

$+C_{5} D_{4} F_{6}-C_{4} D_{5} F_{6})) R_2)) /(A_{3} B_{2} E_{1}-A_{2} B_{3} E_{1}-A_{3} B_{1} E_{2}+A_{1} B_{3} E_{2}+A_{2} B_{1} E_{3}-A_{1} B_{2} E_{3})^{2}$

$Y_{5}=$

$((C_{6} D_{5} F_{4}-C_{5} D_{6} F_{4}-C_{6} D_{4} F_{5}+C_{4} D_{6} F_{5}+C_{5} D_{4} F_{6}-C_{4} D_{5} F_{6}) (B_{3}((-A_{2} E_{1}+A_{1} E_{2})(D_{6} F_{4}-D_{4} F_{6}) c_{0} +C_{6} d_{1} F_{4}$

$(E_{2} A_{0}-A_{2} E_{0})+C_{4} d_{1} F_{6}(-E_{2} A_{0}+A_{2} E_{0})+C_{6}(A_{2} E_{1}-A_{1} E_{2})(F_{4} d_{0}-D_{4} f_{0})-C_{4}(A_{2} E_{1}-A_{1} E_{2})(F_{6} d_{0}-D_{6} f_{0}))$

$+B_{2}((A_{3} E_{1}-A_{1} E_{3})(D_{6} F_{4}-D_{4} F_{6}) c_{0}+C_{4} d_{1} F_{6}(E_{3} A_{0}-A_{3} E_{6})+C_{6} d_{1} F_{4}(-E_{3} A_{0}+A_{3} E_{6})-C_{6}(A_{3} E_{1}-A_{1} E_{3})$

$(F_{4} d_{0}-D_{4} f_{0})+C_{4}(A_{3} E_{1}-A_{1} E_{3})(F_{6} d_{0}-D_{6} f_{0}))+ (A_{3} E_{2}-A_{2} E_{3})(B_{1}(-D_{6} F_{4}+D_{4} F_{6}) c_{0}-C_{6}(d_{1} F_{4} B_{0} -B_{1} F_{4} d_{0}$

$+B_{1} D_{4} f_{0})+C_{4}(d_{1} F_{6} B_{0}-B_{1} F_{6} d_{0}+B_{1} D_{6} f_{0})))+ (A_{3} B_{2} E_{1}-A_{2} B_{3} E_{1}-A_{3} B_{1} E_{2}+A_{1} B_{3} E_{2}+A_{2} B_{1} E_{3}-A_{1} B_{2} E_{3})$

$(x_{25} C_{6} D_{4}-x_{25} C_{4} D_{6}-x_{17} C_{6} F_{4}+x_{12} D_{6} F_{4}+x_{17} C_{4} F_{6}-x_{12} D_{4} F_{6})(-D_{6} F_{5} C_{6}+D_{5} F_{6} C_{6}+C_{6} F_{5} D_{0}-C_{5} F_{6} D_{0}$

$-C_{6} D_{5} F_{6}+C_{5} D_{6} F_{6}) R_1^{2}+ (x_{23}(C_{6} D_{4}-C_{4} D_{6})(-C_{6} D_{5} F_{4}+C_{5} D_{6} F_{4}+C_{6} D_{4} F_{5}-C_{4} D_{6} F_{5}-C_{5} D_{4} F_{6} +C_{4} D_{5} F_{6})$

$(B_{3} E_{2} A_{0}-B_{2} E_{3} A_{0} -A_{3} E_{2} B_{0}+A_{2} E_{3} B_{0}+A_{3} B_{2} E_{0}-A_{2} B_{3} E_{0})+ x_{11}(D_{6} F_{4}-D_{4} F_{6})(-C_{6} D_{5} F_{4}+C_{5} D_{6} F_{4} +C_{6} D_{4} F_{5}$

$-C_{4} D_{6} F_{5}-C_{5} D_{4} F_{6}+C_{4} D_{5} F_{6})(B_{3} E_{2} A_{0}-B_{2} E_{3} A_{0}-A_{3} E_{2} B_{0}+A_{2} E_{3} B_{0}+A_{3} B_{2} E_{0}-A_{2} B_{3} E_{0})- (A_{3} B_{2} E_{1}-A_{2} B_{3} E_{1}$

$-A_{3} B_{1} E_{2}+A_{1} B_{3} E_{2}+A_{2} B_{1} E_{3}-A_{1} B_{2} E_{3})(x_{26} C_{6} D_{4}-x_{26} C_{4} D_{6}+x_{16} C_{6} F_{4}-x_{13} D_{6} F_{4}-x_{16} C_{4} F_{6}+x_{13} D_{4} F_{6})(-D_{6} F_{5} C_{0}$

$+D_{5} F_{6} C_{0}+C_{6} F_{5} D_{0}-C_{5} F_{6} D_{0}-C_{6} D_{5} F_{0}+C_{5} D_{6} F_{0})) R_{1} R_2+(x_{24} C_{6} D_{4}-x_{24} C_{4} D_{6}+x_{16} D_{6} F_{4}-x_{10} D_{4} F_{6})(-C_{6} D_{5} F_{4}$

$+C_{5} D_{6} F_{4}+C_{6} D_{4} F_{5}-C_{4} D_{6} F_{5}-C_{5} D_{4} F_{6}+C_{4} D_{5} F_{6})(B_{3} E_{2} A_{0}-B_{2} E_{3} A_{0}-A_{3} E_{2} B_{0}+A_{2} E_{3} B_{0}+A_{3} B_{2} E_{0} -A_{2} B_{3} E_{0})$

$ R_2^{2}) / ((A_{3} B_{2} E_{1}-A_{2} B_{3} E_{1}-A_{3} B_{1} E_{2}+A_{1} B_{3} E_{2}+A_{2} B_{1} E_{3}-A_{1} B_{2} E_{3})(C_{6} D_{5} F_{4}-C_{5} D_{6} F_{4}-C_{6} D_{4} F_{5} +C_{4} D_{6} F_{5}$

$+C_{5} D_{4} F_{6}-C_{4} D_{5} F_{6})^{2})$

$Y_{3}=$

$(-(A_{3} B_{2} E_{1}-A_{2} B_{3} E_{1}-A_{3} B_{1} E_{2}+A_{1} B_{3} E_{2}+A_{2} B_{1} E_{3}-A_{1} B_{2} E_{3})(C_{6} D_{5} F_{4}-C_{5} D_{6} F_{4}-C_{6} D_{4} F_{5}+C_{4} D_{6} F_{5}+C_{5} D_{4} F_{6}$

$-C_{4} D_{5} F_{6}) (B_{2} E_{1} a_{0}-B_{1} E_{2} a_{0}-A_{2} E_{1} b_{0}+A_{1} E_{2} b_{0}+A_{2} B_{1} e_{0}-A_{1} B_{2} e_{0})+ (x_{21} A_{2} B_{1} -x_{21} A_{1} B_{2}-x_{9} A_{2} E_{1}+x_{3} B_{2} E_{1}$

$+x_{9} A_{1} E_{2}-x_{3} B_{1} E_{2})(A_{3} B_{2} E_{1}-A_{2} B_{3} E_{1}-A_{3} B_{1} E_{2}+A_{1} B_{3} E_{2}+A_{2} B_{1} E_{3}-A_{1} B_{2} E_{3}) ({D}_{6} {F}_{5}{C}_{0}-{D}_{5}{F}_{6} {C}_{0} -{C}_{6} {F}_{5} {D}_{0}$

$+{C}_{5} {F}_{6} {D}_{0}+{C}_{6} {D}_{5} {F}_{0}-{C}_{5} {D}_{6} {F}_{0}) {R}_1^{2}+ ({C}_{6} {D}_{5} {F}_{4}-{C}_{5} {D}_{6} {F}_{4}-{C}_{6} {D}_{4} {F}_{5}+{C}_{4} {D}_{6} {F}_{5}+{C}_{5} {D}_{4} {F}_{6}-{C}_{4} {D}_{5} {F}_{6}) ({B}_{3} {E}_{2} {A}_{0}$

$-{B}_{2} {E}_{3} {A}_{0}-{A}_{3} {E}_{2} {B}_{0}+{A}_{2} {E}_{3} {B}_{0}+{A}_{3} {B}_{2} {E}_{0}-{A}_{2} {B}_{3} {E}_{0}) {R}_2 (-{x}_{7} {A}_{2} {E}_{1}+{x}_{7} {A}_{1} {E}_{2}+(-{x}_{18} {A}_{2} {B}_{1}+{x}_{18} {A}_{1} {B}_{2}-{x}_{6} {A}_{2} {E}_{1}$

$+{x}_{2} {B}_{2} {E}_{1}+{x}_{6} {A}_{1} {E}_{2}-{x}_{2} {B}_{1} {E}_{2}) {R}_2)+ {R_1}({x}_{5}({A}_{2} {E}_{1}-{A}_{1} {E}_{2})({C}_{6} {D}_{5} {F}_{4}-{C}_{5} {D}_{6} {F}_{4}-{C}_{6} {D}_{4} {F}_{5}+{C}_{4} {D}_{6} {F}_{5}+{C}_{5} {D}_{4} {F}_{6}$

$-{C}_{4} {D}_{5} {F}_{6})(-{B}_{3} {E}_{2} {A}_{0}+{B}_{2} {E}_{3} {A}_{0}+{A}_{3} {E}_{2} {B}_{0}-{A}_{2} {E}_{3} {B}_{0}-{A}_{3} {B}_{2} {E}_{0}+{A}_{2} {B}_{3} {E}_{0})+ (-{x}_{19}({A}_{2} {B}_{1}-{A}_{1} {B}_{2})({C}_{6} {D}_{5} {F}_{4}$

$-{C}_{5} {D}_{6} {F}_{4}-{C}_{6} {D}_{4} {F}_{5}+{C}_{4} {D}_{6} {F}_{5}+{C}_{5} {D}_{4} {F}_{6}-{C}_{4} {D}_{5} {F}_{6})(-{B}_{3} {E}_{2} {A}_{0}+{B}_{2} {E}_{3} {A}_{0}+{A}_{3} {E}_{2} {B}_{0}-{A}_{2} {E}_{3} {B}_{0}-{A}_{3} {B}_{2} {E}_{0}+{A}_{2} {B}_{3} {E}_{0})$

$-x_{1}(B_{2} E_{1}-B_{1} E_{2})(C_{6} D_{5} F_{4}-C_{5} D_{6} F_{4}-C_{6} D_{4} F_{5}+C_{4} D_{6} F_{5}+C_{5} D_{4} F_{6}-C_{4} D_{5} F_{6})(-B_{3} E_{2} A_{0}+B_{2} E_{3} A_{0}+A_{3} E_{2} B_{0}-$

$A_{2} E_{3} B_{0}-A_{3} B_{2} E_{0}+A_{2} B_{3} E_{0})-$
$(x_{20} A_{2} B_{1}-x_{20} A_{1} B_{2}+x_{8} A_{2} E_{1}-x_{4} B_{2} E_{1}-x_{8} A_{1} E_{2}+x_{4} B_{1} E_{2})(A_{3} B_{2} E_{1}-A_{2} B_{3} E_{1}-$

$A_{3} B_{1} E_{2}+A_{1} B_{3} E_{2}+A_{2} B_{1} E_{3}-A_{1} B_{2} E_{3}) (D_{6} F_{5} C_{0}-D_{5} F_{6} C_{0}-C_{6} F_{5} D_{0}+C_{5} F_{6} D_{0}+C_{6} D_{5} F_{0}-C_{5} D_{6} F_{0})) R_2)) /$

$((A_{3} B_{2} E_{1}-A_{2} B_{3} E_{1}-A_{3} B_{1} E_{2}+A_{1} B_{3} E_{2}+A_{2} B_{1} E_{3}-A_{1} B_{2} E_{3})^{2}(C_{6} D_{5} F_{4}-C_{5} D_{6} F_{4}-C_{6} D_{4} F_{5}+C_{4} D_{6} F_{5}+C_{5} D_{4} F_{6}$

$-C_{4} D_{5} F_{6}))$

$Y_{6}=$

$((-C_{6} D_{5} F_{4}+C_{5} D_{6} F_{4}+C_{6} D_{4} F_{5}-C_{4} D_{6} F_{5}-C_{5} D_{4} F_{6}+C_{4} D_{5} F_{6}) (B_{3}((-A_{2} E_{1}+A_{1} E_{2})(D_{5} F_{4}-D_{4} F_{5}) c_{0} +C_{5} d_{1} F_{4}$

$(E_{2} A_{0}-A_{2} E_{0})+C_{4} d_{1} F_{5}(-E_{2} A_{0}+A_{2} E_{0})+C_{5}(A_{2} E_{1}-A_{1} E_{2})(F_{4} d_{0}-D_{4} f_{0})- C_{4}(A_{2} E_{1}-A_{1} E_{2})(F_{5} d_{0}-D_{5} f_{0}))+$

$ B_{2}((A_{3} E_{1}-A_{1} E_{3})(D_{5} F_{4}-D_{4} F_{5}) c_{0}+C_{4} d_{1} F_{5}(E_{3} A_{0}-A_{3} E_{0})+C_{5} d_{1} F_{4}(-E_{3} A_{0}+A_{3} E_{0})-C_{5}(A_{3} E_{1}-A_{1} E_{3})(F_{4} d_{0}-$

$D_{4} f_{0})+C_{4}(A_{3} E_{1}-A_{1} E_{3})(F_{5} d_{0}-D_{5} f_{0}))
(A_{3} E_{2}-A_{2} E_{3}) (B_{1} (-D_{5} F_{4}+D_{4} F_{5} ) c_{0}-C_{5} (d_{1} F_{4} B_{0}-B_{1} F_{4} d_{0}+B_{1} D_{4} f_{0} )+$

$C_{4} (d_{1} F_{5} B_{0}-B_{1} F_{5} d_{0}+B_{1} D_{5} f_{0} ) ) )- (A_{3} B_{2} E_{1}-A_{2} B_{3} E_{1}-A_{3} B_{1} E_{2}+A_{1} B_{3} E_{2}+A_{2} B_{1} E_{3}-A_{1} B_{2} E_{3})(x_{25} C_{5} D_{4}-$

$ x_{25} C_{4} D_{5}-x_{17} C_{5} F_{4}+x_{12} D_{5} F_{4}+x_{17} C_{4} F_{5}-x_{12} D_{4} F_{5})$
$(-D_{6} F_{5} C_{0}+D_{5} F_{6} C_{0}+C_{6} F_{5} D_{0}-C_{5} F_{6} D_{0}-C_{6} D_{5} F_{0}+$

$C_{5} D_{6} F_{0}) R_1^{2}+ (x_{23}(C_{5} D_{4}-C_{4} D_{5})(C_{6} D_{5} F_{4}-C_{5} D_{6} F_{4}-C_{6} D_{4} F_{5}+C_{4} D_{6} F_{5}+C_{5} D_{4} F_{6}-C_{4} D_{5} F_{6})(B_{3} E_{2} A_{0}-B_{2} E_{3} A_{0}-$

$A_{3} E_{2} B_{0}+A_{2} E_{3} B_{0}+A_{3} B_{2} E_{0}-A_{2} B_{3} E_{0})+  x_{11}(D_{5} F_{4}-D_{4} F_{5})(C_{6} D_{5} F_{4}-C_{5} D_{6} F_{4}-C_{6} D_{4} F_{5}+C_{4} D_{6} F_{5}+ C_{5} D_{4} F_{6}-$

$C_{4} D_{5} F_{6})(B_{3} E_{2} A_{0}-B_{2} E_{3} A_{0}-A_{3} E_{2} B_{0}+A_{2} E_{3} B_{0}+A_{3} B_{2} E_{0}-A_{2} B_{3} E_{0})+ (A_{3} B_{2} E_{1}-A_{2} B_{3} E_{1}-A_{3} B_{1} E_{2}+A_{1} B_{3} E_{2}+$

$A_{2} B_{1} E_{3}-A_{1} B_{2} E_{3})(x_{26} C_{5} D_{4}-x_{26} C_{4} D_{5}+x_{16} C_{5} F_{4}-x_{13} D_{5} F_{4}-x_{16} C_{4} F_{5}+x_{13} D_{4} F_{5})$
$.(-D_{6} F_{5} C_{0}+D_{5} F_{6} C_{0}+$

$C_{6} F_{5} D_{0}-C_{5} F_{6} D_{0}-C_{6} D_{5} F_{0}+C_{5} D_{6} F_{0})) R_1 R_2+ (x_{24} C_{5} D_{4}-x_{24} C_{4} D_{5}+x_{10} D_{5} F_{4}-x_{10} D_{4} F_{5})(C_{6} D_{5} F_{4}-C_{5} D_{6} F_{4}-$

$C_{6} D_{4} F_{5}+C_{4} D_{6} F_{5}+C_{5} D_{4} F_{6}-C_{4} D_{5} F_{6}) (B_{3} E_{2} A_{0}-B_{2} E_{3} A_{0}-A_{3} E_{2} B_{0}+A_{2} E_{3} B_{0}+A_{3} B_{2} E_{0}- A_{2} B_{3} E_{0}) R_2^{2}) / $

$((A_{3} B_{2} E_{1}- A_{2} B_{3} E_{1}-A_{3} B_{1} E_{2}+A_{1} B_{3} E_{2}+A_{2} B_{1} E_{3}-A_{1} B_{2} E_{3})(C_{6} D_{5} F_{4}-C_{5} D_{6} F_{4}-C_{6} D_{4} F_{5}+C_{4} D_{6} F_{5}+C_{5} D_{4} F_{6}-$

$C_{4} D_{5} F_{6})^{2})$


\textcolor{Sepia}{\subsection{\sffamily Coefficients}\label{sec:appendixB}}

Here are the constants which appear in  Y's where $S=\frac{{P_{a}}}{3a^{2}}$ and $P=\frac{3{P_{a}}^2}{2a^{2}}$ terms are given as:

    \begin{align*} 
    A_{0}=&\frac{1}{\sqrt{2\mathrm{k}}} \cosh \left[R_{1,k}\left(\eta\right)\right] (P \cos \left[\Phi_{1,k}\left(\eta\right)\right]+\mathrm{k} \sin \left[\Phi_{1,k}\left(\eta\right)]\right)+\left(-P \cos \left[2 \Theta_{1,k}\left(\eta\right)+\Phi_{1,k}\left(\eta\right)\right]+\mathrm{k} \sin \left[2 \Theta_{1,k}\left(\eta\right)\right.\right.\\&\left.\left.+\Phi_{1,k}\left(\eta\right)\right]\right) \sinh \mathrm{R}_{1,k}\left(\eta\right)\\
    a_{0}=&\frac{P}{3 \sqrt{2} \sqrt{\mathrm{k}}}\left(\cos \left[\Phi_{2,k}\left(\eta\right)\right] \cosh \left[\mathrm{R}_{2,k}\left(\eta\right)\right]-\cos \left[2 \Theta_{2,k}\left(\eta\right)+\Phi_{2,k}\left(\eta\right)\right]  \sinh \left[\mathrm{R}_{2,k}\left(\eta\right)\right]\right)\\
    A_{1}=&\frac{1}{\sqrt{2} \sqrt{\mathrm{k}}}\left(-\cos \left[2 \Theta_{1,k}\left(\eta\right)+\Phi_{1,k}\left(\eta\right)\right]  \cosh \left[\mathrm{R}_{1,k}\left(\eta\right)\right]+\cos \left[\Phi_{1,k}\left(\eta\right)\right]  \sinh \left[\mathrm{R}_{1,k}\left(\eta\right)\right]\right)\\
    A_{2}=&\frac{1}{\sqrt{\mathrm{k}}}\left(\sqrt{2} \sin \left[2 \Theta_{1,k}\left(\eta\right)+\Phi_{1,k}\left(\eta\right)\right]  \sinh \left[\mathrm{R}_{1,k}\left(\eta\right)\right]\right)\\
    A_{3}=&\frac{1}{\sqrt{2} \sqrt{\mathrm{k}}}\left(-\cosh \left[R_{1,k}\left(\eta\right)\right]  \sin \left[\Phi_{1,k}\left(\eta\right)\right]+\sin \left[2 \Theta_{1,k}\left(\eta\right)+\Phi_{1,k}\left(\eta\right)\right]  \sinh \left[R_{1,k}\left(\eta\right)\right]\right)\\
    B_{0}=&-\frac{1}{\sqrt{2} \sqrt{\mathrm{k}}}\left(\cosh \left[R_{1,k}\left(\eta\right)\right]\left(-\mathrm{k} \cos \left[\Phi_{1,k}\left(\eta\right)\right]+P \sin \left[\Phi_{1,k}\left(\eta\right)\right]\right)-\left(\mathrm{k} \cos \left[2 \Theta_{1,k}\left(\eta\right)+\Phi_{1,k}\left(\eta\right)\right]\right.\right.\\
     &\left.\left.+P \sin \left[2 \Theta_{1,k}\left(\eta\right)+\Phi_{1,k}\left(\eta\right)\right]\right) \sinh \left[R_{1,k}\left(\eta\right)\right]\right)\\
     b_{0}=&\frac{1}{3 \sqrt{2} \sqrt{\mathrm{k}}}\left(P\left(\cosh \left[R_{2,k}\left(\eta\right)\right]  \sin \left[\Phi_{2,k}\left(\eta\right)\right]-\sin \left[2 \Theta_{2,k}\left(\eta\right)+\Phi_{2,k}\left(\eta\right)\right]  \sinh \left[R_{2,k}\left(\eta\right)\right]\right)\right)\\
  B_{1}=&\frac{1}{\sqrt{2} \sqrt{k}}\left(-\cosh \left[R_{1,k}\left(\eta\right)\right]  \sin \left[2 \Theta_{1,k}\left(\eta\right)+\Phi_{1,k}\left(\eta\right)\right]+\sin \left[\Phi_{1,k}\left(\eta\right)\right]  \sinh \left[R_{1,k}\left(\eta\right)\right]\right)\\
  B_{2}=&\frac{-1}{\sqrt{k}}\left(\sqrt{2} \cos \left[2 \Theta_{1,k}\left(\eta\right)+\Phi_{1,k}\left(\eta\right)\right]  \sinh \left[R_{1,k}\left(\eta\right)\right]\right)\\
B_{3}=&\frac{1}{\sqrt{2} \sqrt{k}}\left(\cos \left[\Phi_{1,k}\left(\eta\right)\right]  \cosh \left[R_{1,k}\left(\eta\right)\right]-\cos \left[2 \Theta_{1,k}\left(\eta\right)+\Phi_{1,k}\left(\eta\right)\right]  \sinh \left[R_{1,k}\left(\eta\right)\right]\right)\\ 
\mathrm{C}_{0}=&\frac{1}{\sqrt{2} \sqrt{\mathrm{k}}}\left(\cosh \left[\mathrm{R}_{2,k}\left(\eta\right)\right]\left(P \cos \left[\Phi_{2,k}\left(\eta\right)\right]+\mathrm{k} \sin \left[\Phi_{2,k}\left(\eta\right)\right]\right)+\left(-P\cos \left[2 \Theta_{2,k}\left(\eta\right)+\Phi_{2,k}\left(\eta\right)\right]\right.\right.\\
&\left.\left.+\mathrm{k} \sin \left[2 \Theta_{2,k}\left(\eta\right)+\Phi_{2,k}\left(\eta\right)\right]\right) \sinh \left[\mathrm{R}_{2,k}\left(\eta\right)\right]\right)\\
\mathrm{c}_{0}=&\frac{P}{3 \sqrt{2} \sqrt{\mathrm{k}}}\left(\cos \left[\Phi_{1,k}\left(\eta\right)\right]  \cosh \left[\mathrm{R}_{1,k}\left(\eta\right)\right]-\cos \left[2 \Theta_{1,k}\left(\eta\right)+\Phi_{1,k}\left(\eta\right)\right]  \sinh \left[\mathrm{R}_{1,k}\left(\eta\right)\right]\right)\\
  \mathrm{C}_{4}=&\frac{1}{\sqrt{2} \sqrt{\mathrm{k}}}\left(-\cos \left[2 \Theta_{2,k}\left(\eta\right)+\Phi_{2,k}\left(\eta\right)\right]  \cosh \left[\mathrm{R}_{2,k}\left(\eta\right)\right]+\cos \left[\Phi_{2,k}\left(\eta\right)\right]  \sinh \left[\mathrm{R}_{2,k}\left(\eta\right)\right]\right)\\
  \mathrm{C}_{5}=&\frac{1}{\sqrt{\mathrm{k}}}\left(\sqrt{2} \sin \left[2 \Theta_{2,k}\left(\eta\right)+\Phi_{2,k}\left(\eta\right)\right]  \sinh \left[\mathrm{R}_{2,k}\left(\eta\right)\right]\right)\\
\mathrm{C}_{6}=&\frac{1}{\sqrt{2} \sqrt{\mathrm{k}}}\left(-\cosh \left[\mathrm{R}_{2,k}\left(\eta\right)\right]  \sin \left[\Phi_{2,k}\left(\eta\right)\right]+\sin \left[2 \Theta_{2,k}\left(\eta\right)+\Phi_{2,k}\left(\eta\right)\right]  \sinh \left[\mathrm{R}_{2,k}\left(\eta\right)\right]\right)\\
        {D}_{0}=&\frac{1}{\sqrt{2} \sqrt{\mathrm{k}}}\left(\cosh \left[\mathrm{R}_{2,k}\left(\eta\right)\right]\left(-\mathrm{k} \cos \left[\Phi_{2,k}\left(\eta\right)\right]+P\sin \left[\Phi_{2,k}\left(\eta\right)\right]\right)-\left(\mathrm{k} \cos \left[2 \Theta_{2,k}\left(\eta\right)+\Phi_{2,k}\left(\eta\right)\right]\right.\right.\\
        &\left.\left.+P\sin \left[2 \Theta_{2,k}\left(\eta\right)+\Phi_{2,k}\left(\eta\right)\right]\right) \sinh \left[\mathrm{R}_{2,k}\left(\eta\right)\right]\right)\\
         \mathrm{d}_{0}=&\frac{1}{3 \sqrt{2} \sqrt{\mathrm{k}}}\left(P\left(\cosh \left[\mathrm{R}_{1,k}\left(\eta\right)\right]  \sin \left[\Phi_{1,k}\left(\eta\right)\right]-\sin \left[2 \Theta_{1,k}\left(\eta\right)+\Phi_{1,k}\left(\eta\right)\right]  \sinh \left[\mathrm{R}_{1,k}\left(\eta\right)\right]\right)\right)\\
         \mathrm{D}_{4}=&\frac{1}{\sqrt{2} \sqrt{\mathrm{k}}}\left(-\cosh \left[\mathrm{R}_{2,k}\left(\eta\right)\right]  \sin \left[2 \Theta_{2,k}\left(\eta\right)+\Phi_{2,k}\left(\eta\right)\right]+\sin \left[\Phi_{2,k}\left(\eta\right)\right]  \sinh \left[\mathrm{R}_{2,k}\left(\eta\right)\right]\right) \\
 \mathrm{D}_{5}=&\frac{-1}{\sqrt{\mathrm{k}}}\left(\sqrt{2} \cos \left[2 \Theta_{2,k}\left(\eta\right)+\Phi_{2,k}\left(\eta\right)\right]  \sinh \left[\mathrm{R}_{2,k}\left(\eta\right)\right]\right)\\
 \mathrm{D}_{6}=&\frac{1}{\sqrt{2} \sqrt{\mathrm{k}}}\left(\cos \left[\Phi_{2,k}\left(\eta\right)\right]  \cosh \left[\mathrm{R}_{2,k}\left(\eta\right)\right]-\cos \left[2 \Theta_{2,k}\left(\eta\right)+\Phi_{2,k}\left(\eta\right)\right]  \sinh \left[\mathrm{R}_{2,k}\left(\eta\right)\right]\right)\\
\mathrm{E}_{0}=&\frac{e^{i \Phi_{1,k}\left(\eta\right)}}{\sqrt{2}} \sqrt{\mathrm{k}}\left(\cosh \left[\mathrm{R}_{1,k}\left(\eta\right)\right]\left(-i P+k c_{s 1}^{2}\right)-e^{2 i \Theta_{1,k}\left(\eta\right)} \sinh \left[\mathrm{R}_{1,k}\left(\eta\right)\right]\left(i P+k c_{s1}^{2}\right)\right)\\
    \mathrm{e}_{0}=&\frac{e^{i \Phi_{2,k}\left(\eta\right)}}{3 \sqrt{2} \sqrt{k}}\left((-i k P+3 \mathrm{~S}) \cosh \left[\mathrm{R}_{2,k}\left(\eta\right)\right]+e^{2 i \Theta_{2,k}\left(\eta\right)} i(-k P+3 i \mathrm{~S}) \sinh \left[\mathrm{R}_{2,k}\left(\eta\right)\right]\right)\\
    \mathrm{E}_{1}=&\frac{e^{i \Phi_{1,k}\left(\eta\right)}}{\sqrt{2}} i \sqrt{\mathrm{k}}\left(e^{2 i \Theta_{1,k}\left(\eta\right)} \cosh \left[\mathrm{R}_{1,k}\left(\eta\right)\right]+\sinh \left[\mathrm{R}_{1,k}\left(\eta\right)\right]\right) \\
    \mathrm{E}_{2}=&-\sqrt{2} e^{i\left(2 \Theta_{1,k}\left(\eta\right)+\Phi_{1,k}\left(\eta\right)\right)} \sqrt{\mathrm{k}} \sinh \left[\mathrm{R}_{1,k}\left(\eta\right)\right] \\
\mathrm{E}_{3}=&-\frac{1}{\sqrt{2}} e^{i\left(\Theta_{1,k}\left(\eta\right)+\Phi_{1,k}\left(\eta\right)\right)} \sqrt{\mathrm{k}}\left(\cosh \left[\mathrm{R}_{1,k}\left(\eta\right)+i \Theta_{1,k}\left(\eta\right)\right]+\sinh \left[\mathrm{R}_{1,k}\left(\eta\right)-i \Theta_{1,k}\left(\eta\right)\right]\right)\\
f_{0}=&\frac{e^{i \Phi_{1,k}\left(\eta\right)}}{3 \sqrt{2} \sqrt{k}}\left((-i k P+3 S) \cosh \left[R_{1,k}\left(\eta\right)\right]+e^{2 i \Theta_{1,k}\left(\eta\right)} i(-k P+3 i S) \sinh \left[R_{1,k}\left(\eta\right)\right]\right) \\
F_{0}=&\frac{-e^{i \Phi_{2,k}\left(\eta\right)}}{\sqrt{2}} \sqrt{k}\left(e^{2 i \Theta_{2,k}\left(\eta\right)} \sinh \left[R_{2,k}\left(\eta\right)\right]\left(i P+k c_{s 2}^{2}\right)+i \cosh \left[R_{2,k}\left(\eta\right)\right]\left(P+i k c_{s 2}^{2}\right)\right) \\
F_{4}=&\frac{e^{i \Phi_{2,k}\left(\eta\right)}}{\sqrt{2}} i \sqrt{k}\left(e^{2 i \Theta_{2,k}\left(\eta\right)} \cosh \left[R_{2,k}\left(\eta\right)\right]+\sinh \left[R_{2,k}\left(\eta\right)\right]\right) \\
F_{5}=&-\sqrt{2} e^{i\left(2 \Theta_{2,k}\left(\eta\right)+\Phi_{2,k}\left(\eta\right)\right)} \sqrt{k} \sinh \left[R_{2,k}\left(\eta\right)\right]\\
F_{6}=&-\frac{1}{\sqrt{2}} e^{i \Phi_{2,k}\left(\eta\right)} \sqrt{k}\left(\cosh \left[R_{2,k}\left(\eta\right)\right]+e^{2 i \Theta_{2,k}\left(\eta\right)} \sinh \left[R_{2,k}\left(\eta\right)\right]\right)\\
    x_{1}=&\frac{1}{6 \sqrt{2\mathrm{k}}} \sin\left[2 \Theta_{1,k}\left(\eta\right)\right]\left(-2 \cos \left[\Phi_{1,k}\left(\eta\right)\right]  \cosh \left[R_{1,k}\left(\eta\right)\right]  \sin \left[2\left(\Theta_{1,k}\left(\eta\right)+\Theta_{2,k}\left(\eta\right)\right)\right]\right.\\
    &+\left(\sin \left[2 \Theta_{2,k}\left(\eta\right)-\Phi_{1,k}\left(\eta\right)\right]+2 \sin \left[4 \Theta_{1,k}\left(\eta\right)-2\Theta_{2,k}\left(\eta\right)+\Phi_{1,k}\left(\eta\right)\right]-\sin \left[4 \Theta_{1,k}\left(\eta\right)+2 \Theta_{2,k}\left(\eta\right)\right.\right.\\
    &\left.\left.\left.+\Phi_{1,k}\left(\eta\right)\right]\right) \sinh \left[R_{1,k}\left(\eta\right)\right]\right)\\
    x_{2}=&\frac{1}{12 \sqrt{2\mathrm{k}}} \sin \left[2 \Theta_{2,k}\left(\eta\right)\right]\left(2 \cos \left[\Phi_{1,k}\left(\eta\right)\right]  \cosh \left[R_{1,k}\left(\eta\right)\right]  \sin \left[2\left(\Theta_{1,k}\left(\eta\right)+\Theta_{2,k}\left(\eta\right)\right)\right]\right.\\
&+\left(\sin \left[2 \Theta_{2,k}\left(\eta\right)-\Phi_{1,k}\left(\eta\right)\right]+2 \sin \left[4 \Theta_{1,k}\left(\eta\right)-2 \Theta_{2,k}\left(\eta\right)+\Phi_{1,k}\left(\eta\right)\right]\right.\\
&\left.\left.-\sin \left[4 \Theta_{1,k}\left(\eta\right)+2 \Theta_{2,k}\left(\eta\right)+\Phi_{1,k}\left(\eta\right)\right]\right) \sinh \left[R_{1,k}\left(\eta\right)\right]\right)\\
   x_{3}=&-\frac{1}{12 \sqrt{2} \sqrt{k}} \sin \left[2 \Theta_{1,k}\left(\eta\right)\right]\left(2 \cos \left[\Phi_{1,k}\left(\eta\right)\right]  \cosh \left[R_{1,k}\left(\eta\right)\right]  \sin \left[2\left(\Theta_{1,k}\left(\eta\right)+\Theta_{2,k}\left(\eta\right)\right)\right]\right.\\
     &+\left(\sin \left[2 \Theta_{2,k}\left(\eta\right)-\Phi_{1,k}\left(\eta\right)\right]+2 \sin \left[4 \Theta_{1,k}\left(\eta\right)-2 \Theta_{2,k}\left(\eta\right)+\Phi_{1,k}\left(\eta\right)\right]-\sin \left[4 \Theta_{1,k}\left(\eta\right)+2 \Theta_{2,k}\left(\eta\right)\right.\right.\\
     &\left.\left.\left.
     +\Phi_{1,k}\left(\eta\right)\right]\right) \sinh \left[R_{1,k}\left(\eta\right)\right]\right)\\
     x_{4}=&\frac{1}{6 \sqrt{2 k}} \sin \left[2 \Theta_{2,k}\left(\eta\right)\right]\left(2 \cos \left[\Phi_{1,k}\left(\eta\right)\right]  \cosh \left[R_{1,k}\left(\eta\right)\right]  \sin \left[2\left(\Theta_{1,k}\left(\eta\right)+\Theta_{2,k}\left(\eta\right)\right)\right]\right.\\
 &+\left(\sin \left[2 \Theta_{2,k}\left(\eta\right)-\Phi_{1,k}\left(\eta\right)\right]+2 \sin \left[4 \Theta_{1,k}\left(\eta\right)-2 \Theta_{2,k}\left(\eta\right)+\Phi_{1,k}\left(\eta\right)\right]-\sin \left[4 \Theta_{1,k}\left(\eta\right)+2 \Theta_{2,k}\left(\eta\right)\right.\right.\\
 &\left.\left.\left.+\Phi_{1,k}\left(\eta\right)\right]\right) \sinh \left[R_{1,k}\left(\eta\right)\right]\right)\\
    x_{6}=&\frac{1}{6 \sqrt{2} \sqrt{k}}\left(\cosh \left[R_{1,k}\left(\eta\right)\right]  \sin \left[2 \Theta_{2,k}\left(\eta\right)\right]  \sin \left[2\left(\Theta_{1,k}\left(\eta\right)+\Theta_{2,k}\left(\eta\right)\right)\right]  \sin [\Phi_{1,k}\left(\eta\right)]\right)\\
    x_{7}=&\left(\cos \left[2\Theta_{2,k}\left(\eta\right)-\Phi_{1,k}\left(\eta\right)\right]-2 \cos \left[4 \Theta_{1,k}\left(\eta\right)-2 \Theta_{2,k}\left(\eta\right)+\Phi_{1,k}\left(\eta\right)\right]+\cos \left[4 \Theta_{1,k}\left(\eta\right)+2 \Theta_{2,k}\left(\eta\right)\right.\right.\\
    &\left.\left.+\Phi_{1,k}\left(\eta\right)\right]\right) \sin \left[2 \Theta_{2,k}\left(\eta\right)\right]\\
    x_{8}=&\frac{1}{6 \sqrt{2} \sqrt{\mathrm{k}}} \sin \left[2 \Theta_{2,k}\left(\eta\right)\right]
\left(\cosh \left[R_{1,k}\left(\eta\right)\right]\left(-\cosh \left[i\left(2\left(\Theta_{1,k}\left(\eta\right)+\Theta_{2,k}\left(\eta\right)\right)+\Phi_{1,k}\left(\eta\right)\right)\right]\right.\right.\\
&\left.+\cosh \left[2 i\left(\Theta_{1,k}\left(\eta\right)+\Theta_{2,k}\left(\eta\right)\right)-i \Phi_{1,k}\left(\eta\right)\right]\right)+\left(\cos \left[2 \Theta_{2,k}\left(\eta\right)-\Phi_{1,k}\left(\eta\right)\right]-2 \cos \left[4 \Theta_{1,k}\left(\eta\right)-2 \Theta_{2,k}\left(\eta\right)+\Phi_{1,k}\left(\eta\right)\right]\right.\\
&\left.\left.+\cos \left[4 \Theta_{1,k}\left(\eta\right)+2 \Theta_{2,k}\left(\eta\right)+\Phi_{1,k}\left(\eta\right)\right]\right) \sinh \left[R_{1,k}\left(\eta\right)\right]\right)\\
x_{9}=&\frac{1}{12 \sqrt{2} \sqrt{\mathrm{k}}} \sin \left[2 \Theta_{1,k}\left(\eta\right)\right]
\left(\cosh \left[R_{1,k}\left(\eta\right)\right]\left(\cosh \left[i\left(2\left(\Theta_{1,k}\left(\eta\right)+\Theta_{2,k}\left(\eta\right)\right)+\Phi_{1,k}\left(\eta\right)\right)\right]-\cosh \left[2 i\left(\Theta_{1,k}\left(\eta\right)+\Theta_{2,k}\left(\eta\right)\right)\right.\right.\right.\\
&\left.\left.-i \Phi_{1,k}\left(\eta\right)\right]\right)-\left(\cos \left[2 \Theta_{2,k}\left(\eta\right)-\Phi_{1,k}\left(\eta\right)\right]-2 \cos \left[4 \Theta_{1,k}\left(\eta\right)-2 \Theta_{2,k}\left(\eta\right)+\Phi_{1,k}\left(\eta\right)\right]\right.\\
&\left.\left.+\cos \left[4 \Theta_{1,k}\left(\eta\right)+2 \Theta_{2,k}\left(\eta\right)+\bar{\Phi}_{1,k}\left(\eta\right)\right]\right) \sinh \left[R_{1,k}\left(\eta\right)\right]\right)\\
x_{10}=&\frac{1}{6 \sqrt{2}(\sqrt{k})} \sin \left[2 \Theta_{2,k}\left(\eta\right)\right]\left(\cosh \left[\mathrm{R}_{2,k}\left(\eta\right)\right]\left(\cos \left[\Phi_{2,k}\left(\eta\right)\right]  \sin \left[2\left(\Theta_{1,k}\left(\eta\right)-\Theta_{2,k}\left(\eta\right)\right)\right]\right.\right.\\
&\left.\left.-2 \sin \left[2 \Theta_{1,k}\left(\eta\right)\right]  \sin \left[2 \Theta_{2,k}\left(\eta\right)\right]  \sin \left[\Phi_{2,k}\left(\eta\right)\right]\right)+\cos \left[2 \Theta_{2,k}\left(\eta\right)+\Phi_{2,k}\left(\eta\right)\right]  \sin \left[2\left(\Theta_{1,k}\left(\eta\right)+\Theta_{2,k}\left(\eta\right)\right)\right]  \sinh \left[R_{2,k}\left(\eta\right)\right]\right)\\
x_{11}=& \frac{-1}{3 \sqrt{2}\left(\sqrt{k}\right)} \sin \left[2 \Theta_{1,k}\left(\eta\right)\right]\left(\cosh \left[R_{2,k}\left(\eta\right)\right]\left(\cos \left[\Phi_{2,k}\left(\eta\right)\right]  \sin \left[2\left(\Theta_{1,k}\left(\eta\right)-\Theta_{2,k}\left(\eta\right)\right)\right]\right.\right.\\
&\left.\left.-2 \sin \left[2 \Theta_{1,k}\left(\eta\right)\right]  \sin \left[2 \Theta_{2,k}\left(\eta\right)\right]  \sin \left[\Phi_{2,k}\left(\eta\right)\right]\right)+\cos \left[2 \Theta_{2,k}\left(\eta\right)+\Phi_{2,k}\left(\eta\right)\right]  \sin \left[2\left(\Theta_{1,k}\left(\eta\right)+\Theta_{2,k}\left(\eta\right)\right)\right]  \sinh \left[R_{2,k}\left(\eta\right)\right]\right)\\
x_{12}=&-\frac{1}{6 \sqrt{2} \sqrt{k}} \sin \left[2 \Theta_{1,k}\left(\eta\right)\right] \left(\cosh \left[R_{2,k}\left(\eta\right)\right]\left(\cos \left[\Phi_{2,k}\left(\eta\right)\right]  \sin \left[2\left(\Theta_{1,k}\left(\eta\right)-\Theta_{2,k}\left(\eta\right)\right)\right]\right.\right.\\
&\left.\left.-2 \sin \left[2 \Theta_{1,k}\left(\eta\right)\right]  \sin \left[2 \Theta_{2,k}\left(\eta\right)\right]  \sin \left[\Phi_{2,k}\left(\eta\right)\right]\right)+\cos \left[2 \Theta_{2,k}\left(\eta\right)+\Phi_{2,k}\right]  \sin \left[2\left(\Theta_{1,k}\left(\eta\right)+\Theta_{2,k}\left(\eta\right)\right)\right] \sinh \left[R_{2,k}\left(\eta\right)\right]\right)\\
 x_{13}=& \frac{1}{3 \sqrt{2} \sqrt{k}} \sin \left[2 \Theta_{2,k}\left(\eta\right)\right] (\cosh \left[R_{2,k}\left(\eta\right)\right](\cos [\Phi_{2,k}\left(\eta\right)]  \sin [2(\Theta_{1,k}\left(\eta\right)-\Theta_{2,k}\left(\eta\right))]\\
 &\left.\left.-2 \sin \left[2 \Theta_{1,k}\left(\eta\right)\right]  \sin \left[2 \Theta_{2,k}\left(\eta\right)\right] \sin \left[\Phi_{2,k}\left(\eta\right)\right]\right)+\cos \left[2 \Theta_{2,k}\left(\eta\right)+\Phi_{2,k}\left(\eta\right)\right]  \sin \left[2\left(\Theta_{1,k}\left(\eta\right)+\Theta_{2,k}\left(\eta\right)\right)\right]  \sinh \left[R_{2,k}\left(\eta\right)\right]\right)\\
  x_{14}=&-\frac{1}{3 \sqrt{2k}} \sin \left[2 \Theta_{1,k}\left(\eta\right)\right] \left(\cosh \left[R_{2,k}\left(\eta\right)\right]\left(2 \cos \left[\Phi_{2,k}\left(\eta\right)\right]  \sin \left[2 \Theta_{1,k}\left(\eta\right)\right]  \sin \left[2 \Theta_{2,k}\left(\eta\right)\right]+\sin \left[2\left(\Theta_{1,k}\left(\eta\right)\right.\right.\right.\right.\\
  &\left.\left.\left.\left.-\Theta_{2,k}\left(\eta\right)\right)\right]  \sin \left[\Phi_{2,k}\left(\eta\right)\right]\right)+\sin \left[2\left(\Theta_{1,k}\left(\eta\right)+\Theta_{2,k}\left(\eta\right)\right)\right]  \sin \left[2 \Theta_{2,k}\left(\eta\right)+\Phi_{2,k}\left(\eta\right)\right] \sinh \left[R_{2,k}\left(\eta\right)\right]\right)\\
  x_{15}=& \frac{1}{6 \sqrt{2} \sqrt{k}} \sin \left[2 \Theta_{2,k}\left(\eta\right)\right] \left(\cosh \left[R_{2,k}\left(\eta\right)\right]\left(2 \cos \left[\Phi_{2,k}\left(\eta\right)\right]  \sin \left[2 \Theta_{1,k}\left(\eta\right)\right]  \sin \left[2 \Theta_{2,k}\left(\eta\right)\right]\right.\right.\\ &\left.\left.+\sin \left[2\left(\Theta_{1,k}\left(\eta\right)-\Theta_{2,k}\left(\eta\right)\right)\right]  \sin \left[\Phi_{2,k}\left(\eta\right)\right]\right)+\sin \left[2\left(\Theta_{1,k}\left(\eta\right)+\Theta_{2,k}\left(\eta\right)\right)\right]  \sin \left[2 \Theta_{2,k}\left(\eta\right)+\Phi_{2,k}\left(\eta\right)\right]  \sinh \left[R_{2,k}\left(\eta\right)\right]\right)\\
  x_{16}=&\frac{1}{12 \sqrt{2}} \mathrm{i} \mathrm{e}^{\mathrm{i}(\Theta_{1,k}\left(\eta\right)+\Phi_{1,k}\left(\eta\right))} \sqrt{\mathrm{k}} \sin \left[2 \Theta_{1,k}\left(\eta\right)\right] \left({e}^{-\mathrm{i} \Theta_{1,k}\left(\eta\right)} \sin \left[2 \Theta_{2,k}\left(\eta\right)\right]\left(2 \cos \left[2 \Theta_{1,k}\left(\eta\right)\right]  \cosh \left[\mathrm{R}_{1,k}\left(\eta\right)\right]\right.\right.\\
  &\left.-\left(1-3 e^{4 i \Theta_{1,k}\left(\eta\right)}\right) \sinh \left[\mathrm{R}_{1,k}\left(\eta\right)\right]\right)+  2 \cos \left[2 \Theta_{2,k}\left(\eta\right)\right]  \sin \left[2 \Theta_{1,k}\left(\eta\right)\right]\left(\cosh \left[\mathrm{R}_{1,k}\left(\eta\right)-\mathrm{i} \Theta_{1,k}\left(\eta\right)\right]\right.\\
  &\left.\left.-\sinh \left[\mathrm{R}_{1,k}\left(\eta\right)+i \Theta_{1,k}\left(\eta\right)\right]\right)\right)\\
   x_{17}=&-\frac{1}{6 \sqrt{2}} i e^{i\left(\Theta_{1,k}\left(\eta\right)+\Phi_{1,k}\left(\eta\right)\right)} \sqrt{k} \sin \left[2 \Theta_{2,k}\left(\eta\right)\right] \left(e^{-i \Theta_{1,k}\left(\eta\right)} \sin \left[2 \Theta_{2,k}\left(\eta\right)\right]\left(2 \cos \left[2 \Theta_{1,k}\left(\eta\right)\right] \cosh \left[R_{1,k}\left(\eta\right)\right]\right.\right.\\&\left.-\left(1-3 e^{4 i \Theta_{1,k}\left(\eta\right)}\right) \sinh \left[R_{1,k}\left(\eta\right)\right]\right)+ 2 \cos \left[2 \Theta_{2,k}\left(\eta\right)\right]  \sin \left[2 \Theta_{1,k}\left(\eta\right)\right]\left(\cosh \left[R_{1,k}\left(\eta\right)-i \Theta_{1,k}\left(\eta\right)\right]\right.\\
   &\left.\left.-\sinh \left[R_{1,k}\left(\eta\right)+i \Theta_{1,k}\left(\eta\right)\right]\right)\right)\\
   x_{18}=& \frac{1}{6 \sqrt{2}} e^{i \Phi_{1,k}\left(\eta\right)} i \sqrt{\mathrm{k}} \sin \left[2 \Theta_{1,k}\left(\eta\right)\right] \left(\sin \left[2 \Theta_{2,k}\left(\eta\right)\right]\left(2 \cos \left[2 \Theta_{1,k}\left(\eta\right)\right]  \cosh \left[R_{1,k}\left(\eta\right)\right]-\left(1-3 e^{4 i \Theta_{1,k}\left(\eta\right)}\right) \sinh \left[R_{1,k}\left(\eta\right)\right]\right)\right.\\ 
   &\left.+2 e^{i \Theta_{1,k}\left(\eta\right)} \cos \left[2 \Theta_{2,k}\left(\eta\right)\right]  \sin \left[2 \Theta_{1,k}\left(\eta\right)\right]\left(\cosh \left[R_{1,k}\left(\eta\right)-i \Theta_{1,k}\left(\eta\right)\right]-\sinh \left[R_{1,k}\left(\eta\right)+i \Theta_{1,k}\left(\eta\right)\right]\right)\right)\\
  x_{19}=& \frac{1}{12 \sqrt{2}} e^{i \Phi_{1,k}\left(\eta\right)} i \sqrt{k} \sin \left[2 \Theta_{2,k}\left(\eta\right)\right]  \left(\sin \left[2 \Theta_{2,k}\left(\eta\right)\right]\left(2 \cos \left[2 \Theta_{1,k}\left(\eta\right)\right]  \cosh \left[R_{1,k}\left(\eta\right)\right]-\left(1-3 e^{4 i \Theta_{1,k}\left(\eta\right)}\right) \sinh \left[R_{1,k}\left(\eta\right)\right]\right)\right.\\
  &\left.+2 e^{i \Theta_{1,k}\left(\eta\right)} \cos \left[2 \Theta_{2,k}\left(\eta\right)\right]  \sin \left[2 \Theta_{1,k}\left(\eta\right)\right]\left(\cosh \left[R_{1,k}\left(\eta\right)-i \Theta_{1,k}\left(\eta\right)\right]-\sinh \left[R_{1,k}\left(\eta\right)+i \Theta_{1,k}\left(\eta\right)\right]\right)\right)\\
  x_{20}=& \frac{1}{6 \sqrt{2}} i e^{i\left(\Theta_{1,k}\left(\eta\right)+\Phi_{1,k}\left(\eta\right)\right)} \sqrt{k} \sin \left[2 \Theta_{2,k}\left(\eta\right)\right]  \left(e^{-i \Theta_{1,k}\left(\eta\right)} \sin \left[2 \Theta_{2,k}\left(\eta\right)\right]\left(2 \cos \left[2 \Theta_{1,k}\left(\eta\right)\right]  \cosh \left[R_{1,k}\left(\eta\right)\right]\right.\right.\\
  &\left.-\left(1-3 e^{4 i \Theta_{1,k}\left(\eta\right)}\right) \sinh \left[R_{1,k}\left(\eta\right)\right]\right)+2 \cos \left[2 \Theta_{2,k}\left(\eta\right)\right]  \sin \left[2 \Theta_{1,k}\left(\eta\right)\right]\left(\cosh \left[R_{1,k}\left(\eta\right)-i \Theta_{1,k}\left(\eta\right)\right]\right.\\&\left.\left.-\sinh \left[R_{1,k}\left(\eta\right)+i \Theta_{1,k}\left(\eta\right)\right]\right)\right)\\
  x_{21}=& \frac{1}{12 \sqrt{2}} i e^{i\left(\Theta_{1,k}\left(\eta\right)+\Phi_{1,k}\left(\eta\right)\right)} \sqrt{k} \sin \left[2 \Theta_{1,k}\left(\eta\right)\right]  \left(e^{-i \Theta_{1,k}\left(\eta\right)} \sin \left[2 \Theta_{2,k}\left(\eta\right)\right]\left(2 \cos \left[2 \Theta_{1,k}\left(\eta\right)\right] \cosh \left[R_{1,k}\left(\eta\right)\right]\right.\right.\\ &\left.-\left(1-3 e^{4 i \Theta_{1,k}\left(\eta\right)}\right) \sinh \left[R_{1,k}\left(\eta\right)\right]\right)+2 \cos \left[2 \Theta_{2,k}\left(\eta\right)\right]  \sin \left[2 \Theta_{1,k}\left(\eta\right)\right]\left(\cosh \left[R_{1,k}\left(\eta\right)-i \Theta_{1,k}\left(\eta\right)\right]\right.\\
  &\left.\left.-\sinh \left[R_{1,k}\left(\eta\right)+i \Theta_{1,k}\left(\eta\right)\right]\right)\right)\\
    x_{23}=& \frac{1}{6 \sqrt{2}} e^{i\left(\Theta_{2,k}\left(\eta\right)+\Phi_{2,k}\left(\eta\right)\right)} \sqrt{k} \sin \left[2 \Theta_{1,k}\left(\eta\right)\right] \\& \left(\cosh \left[R_{2,k}\left(\eta\right)-2 i \Theta_{1,k}\left(\eta\right)-3 i \Theta_{2,k}\left(\eta\right)\right]-\cosh \left[R_{2,k}\left(\eta\right)+2 i \Theta_{1,k}\left(\eta\right)-i \Theta_{2,k}\left(\eta\right)\right]\right.\\&-\cosh \left[R_{2,k}\left(\eta\right)-2 i \Theta_{1,k}\left(\eta\right)+i \Theta_{2,k}\left(\eta\right)\right] +\cosh \left[R_{2,k}\left(\eta\right)-i\left(2 \Theta_{1,k}\left(\eta\right)+\Theta_{2,k}\left(\eta\right)\right)\right]
    \\&\left.+2 i \cosh \left[R_{2,k}\left(\eta\right)\right]  \sin \left[2 \Theta_{1,k}\left(\eta\right)-\Theta_{2,k}\left(\eta\right)\right]-2 i \cosh \left[R_{2,k}\left(\eta\right)+2 i\left(\Theta_{1,k}\left(\eta\right)+\Theta_{2,k}\left(\eta\right)\right)\right]  \sin \left[\Theta_{2,k}\left(\eta\right)\right]\right)\\
    x_{24}=& \frac{1}{12 \sqrt{2}} e^{i\left(\Theta_{2,k}\left(\eta\right)+\Phi_{2,k}\left(\eta\right)\right)} \sqrt{k}\\ &\left(-\cosh \left[R_{2,k}\left(\eta\right)-2 i \Theta_{1,k}\left(\eta\right)-3 i \Theta_{2,k}\left(\eta\right)\right]+\cosh \left[R_{2,k}\left(\eta\right)+2 i \Theta_{1,k}\left(\eta\right)-i \Theta_{2,k}\left(\eta\right)\right]\right.\\&+\cosh \left[R_{2,k}\left(\eta\right)-2 i \Theta_{1,k}\left(\eta\right)+i \Theta_{2,k}\left(\eta\right)\right]- \cosh \left[R_{2,k}\left(\eta\right)-i\left(2 \Theta_{1,k}\left(\eta\right)+\Theta_{2,k}\left(\eta\right)\right)\right]\\&-2 i \cosh \left[R_{2,k}\left(\eta\right)\right]  \sin \left[2 \Theta_{1,k}\left(\eta\right)-\Theta_{2,k}\left(\eta\right)\right]+2 i \cosh \left[R_{2,k}\left(\eta\right)\right.\\&\left.\left.+2 i\left(\Theta_{1,k}\left(\eta\right)+\Theta_{2,k}\left(\eta\right)\right)\right]  \sin \left[\Theta_{2,k}\left(\eta\right)\right]\right) \sin \left[2 \Theta_{2,k}\left(\eta\right)\right]\\
    x_{25}=& \frac{1}{12 \sqrt{2}} e^{-R_{2,k}\left(\eta\right)+i\left(\Theta_{2,k}\left(\eta\right)+\Phi_{2,k}\left(\eta\right)\right)} \sqrt{k} \sin \left[2 \Theta_{1,k}\left(\eta\right)\right]  \left(-\cos \left[2 \Theta_{1,k}\left(\eta\right)-\Theta_{2,k}\left(\eta\right)\right]+\cos \left[2 \Theta_{1,k}\left(\eta\right)+3 \Theta_{2,k}\left(\eta\right)\right]\right.\\
&+2 i \cos \left[\Theta_{2,k}\left(\eta\right)\right]  \sin \left[2 \Theta_{1,k}\left(\eta\right)\right]-2 e^{2 R_{2,k}\left(\eta\right)}\left(2 i \cos \left[\Theta_{2,k}\left(\eta\right)\right]  \cos \left[2 \Theta_{1,k}\left(\eta\right)+\Theta_{2,k}\left(\eta\right)\right]\right.\\
&\left.\left.+\sin \left[2 \Theta_{1,k}\left(\eta\right)\right]\right) \sin \left[\Theta_{2,k}\left(\eta\right)\right]\right)\\
x_{26}=& \frac{1}{6 \sqrt{2}} e^{-R_{2,k}\left(\eta\right)+i\left(\Theta_{2,k}\left(\eta\right)+\Phi_{2,k}\left(\eta\right)\right)} \sqrt{k}  \left(-\cos \left[2 \Theta_{1,k}\left(\eta\right)-\Theta_{2,k}\left(\eta\right)\right]+\cos \left[2 \Theta_{1,k}\left(\eta\right)+3 \Theta_{2,k}\left(\eta\right)\right]\right.\\ 
&+2 i \cos \left[\Theta_{2,k}\left(\eta\right)\right]  \sin \left[2 \Theta_{1,k}\left(\eta\right)\right]-2 e^{2 R_{2,k}\left(\eta\right)}\left(2 i \cos \left[\Theta_{2,k}\left(\eta\right)\right] \cos \left[2 \Theta_{1,k}\left(\eta\right)+\Theta_{2,k}\left(\eta\right)\right]\right.\\
&\left.\left.+\sin \left[2 \Theta_{1,k}\left(\eta\right)\right]\right) \sin \left[\Theta_{2,k}\left(\eta\right)\right]\right) \sin \left[2 \Theta_{2,k}\left(\eta\right)\right]
\end{align*}

\twocolumngrid

\textbf{Corresponding author address:}\\
E-mail:~sayantan.choudhury@icts.res.in,  \\
$~~~~~~~~~~~~$sayanphysicsisi@gmail.com

\bibliography{referencesnew}
\bibliographystyle{utphys}

\end{document}